\documentclass[10pt,floatfix]{article}
\usepackage{bm}
\usepackage{graphicx}
\usepackage{latexsym}
\usepackage{amsmath}
\usepackage{amsfonts}
\usepackage{amssymb}
\begin{document}

\title{Towards MIGO, the \textit{M}atter-wave \textit{I}nterferometric
  \textit{G}ravitational-wave \textit{O}bservatory, and the
  Intersection of Quantum Mechanics with General Relativity}
\author{Raymond Y.~Chiao and Achilles D.~Speliotopoulos\\University of
  California, Berkeley, CA 94720-7300}
\date{March 8, 2004}
\maketitle

\begin{abstract}
A dynamical, non-Euclidean spacetime geometry in general relativity theory
implies the possibility of gravitational radiation. Here we explore
novel methods of detecting such radiation from astrophysical sources
by means of matter-wave interferometers (MIGOs), using atomic beams
emanating from supersonic atomic sources that are further cooled and
collimated by means of optical molasses. While the sensitivities of
such MIGOs compare favorably with LIGO and LISA, the sizes of MIGOs
can be orders of magnitude smaller, and their bandwidths wider. Using
a pedagogical approach, we place this problem into the broader context
of problems at the intersection of quantum mechanics with general relativity.
\end{abstract}

\section{Introduction}

In this contribution to the proceedings of the 2003 Gargnano workshop on
`Mysteries, Puzzles, and Paradoxes of Quantum Mechanics', let us begin with a
brief review of those basic notions of differential geometry that will
be important for the understanding of the interaction between
gravitational radiation and quantum matter. Specifically, we shall
examine here matter-wave interferometry as a means of detecting
gravitational radiation from astrophysical sources. In a pedagogical
manner, we shall place this problem into the broader context of some of the
fascinating problems at the intersection of quantum mechanics
with general relativity.

\begin{figure}[ptb]
\begin{center}
\includegraphics[width=0.5\textwidth]
{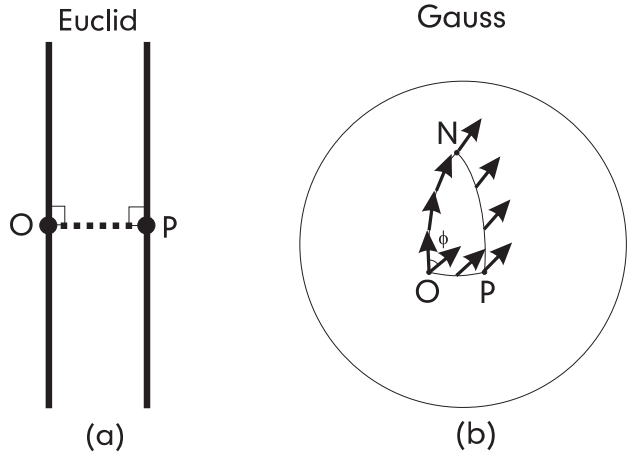}
\end{center}
\caption{(a) Euclid's parallel-lines axiom in plane geometry states that two
straight, parallel lines, such as the ones extended through the nearby
points O and P, never meet at infinity. Gauss discovered a
non-Euclidean geometry which is illustrated in (b). Two parallel
geodesics on a sphere, represented by two lines of longitude emanating
from points O and P on the equator of the sphere, converge onto the
north pole N. Parallel transport of a vector around the spherical
triangle ONP, and back to O again, illustrates the Gauss-Bonnet
theorem, which states that the sum of the three interior angles of a
triangle is 180$^{\circ}$ plus the solid angle subtended by this
triangle with respect to the centre of the sphere. Here $\phi$ is the
rotation angle of the parallel-transported vector after its completion
of the circuit. In Berry's phase, an `anholonomy' is the quantum phase factor
$\exp(im_s\phi)$ that a spin with component $m_s$ normal to the surface would
pick up after adiabatic transport around the circuit OPNO in the field
of a magnetic monopole located at the centre of the sphere.}
\label{euclid-plus-gauss}
\end{figure}

One of Euclid's axioms for plane geometry is the \textit{parallel-lines
axiom}, i.e., that parallel lines never meet, no matter how far
these lines are extended in either direction. This axiom is illustrated by
figure \ref{euclid-plus-gauss}(a). We shall presently extend this axiom of
Euclidean geometry to a statement about worldlines of test particles
in flat spacetimes in general relativity.

Gauss discovered \textit{non-Euclidean} geometry after a
deliberate omission of this Euclidean axiom \cite{SpivakI,
SpivakII}. He found one example of this geometry in the surface of
a sphere, where there exists a constant positive curvature
everywhere. This Gaussian geometry is illustrated in figure
\ref{euclid-plus-gauss}(b). A violation of the Euclidean axiom can
be illustrated by extending northward two initially parallel
`lines of longitude' starting at points O and P at the equator.
These lines of longitude are \textit{geodesics}, i.e., the
`straightest' possible lines between two points, which minimize
the intervening distance between them. When these geodesics are
extended north, they will converge onto, and eventually intersect
at, the north pole N, thus violating Euclid's parallel-line axiom.

Riemann generalized Gauss's notion of non-Euclidean geometry to that of
\textit{differential geometry} for arbitrarily curved manifolds
\cite{SpivakII}. As a measure of local curvature, he introduced the
operational method of the \textit{parallel transport} of a vector around an
infinitesimal circuit, i.e., a tiny closed curve on a differentiable
manifold. After the parallel transport is completed, there results a
rotation of the final direction of the parallel-transported vector by
an angle $\phi$ with respect to its initial direction, as illustrated
in part (b) of figure \ref{euclid-plus-gauss}. This rotation angle $\phi$
is thus a measure of the local curvature of the manifold.

To quantify this angle $\phi$, Riemann introduced his curvature tensor
$R_{ijk}^{l}$, which we define following Landau and Lifschitz's
approach as \cite{LandauCTF}:
\begin{equation}
\delta\xi_{i}=-\frac{1}{2}R_{ijk}^{l}\delta f^{jk}\xi_{l}
\end{equation}
where the Latin indices $i,j,k,l$, which run from 1 to 2 to 3,
represent the three spatial dimensions, $\xi_{l}$ are components of
the initial vector, and $\delta\xi_{i}$ are the changes of the final
vector components relative to the initial components, after the
process of parallel transport of the vector around a closed
circuit\textemdash with a infinitesimal area given by $\delta
f^{jk}=\delta x^j\delta x^k$\textemdash is completed (Einstein's
summation convention is used here for repeated indices).

\begin{figure}[ptb]
\begin{center}
\includegraphics[width=0.99\textwidth]
{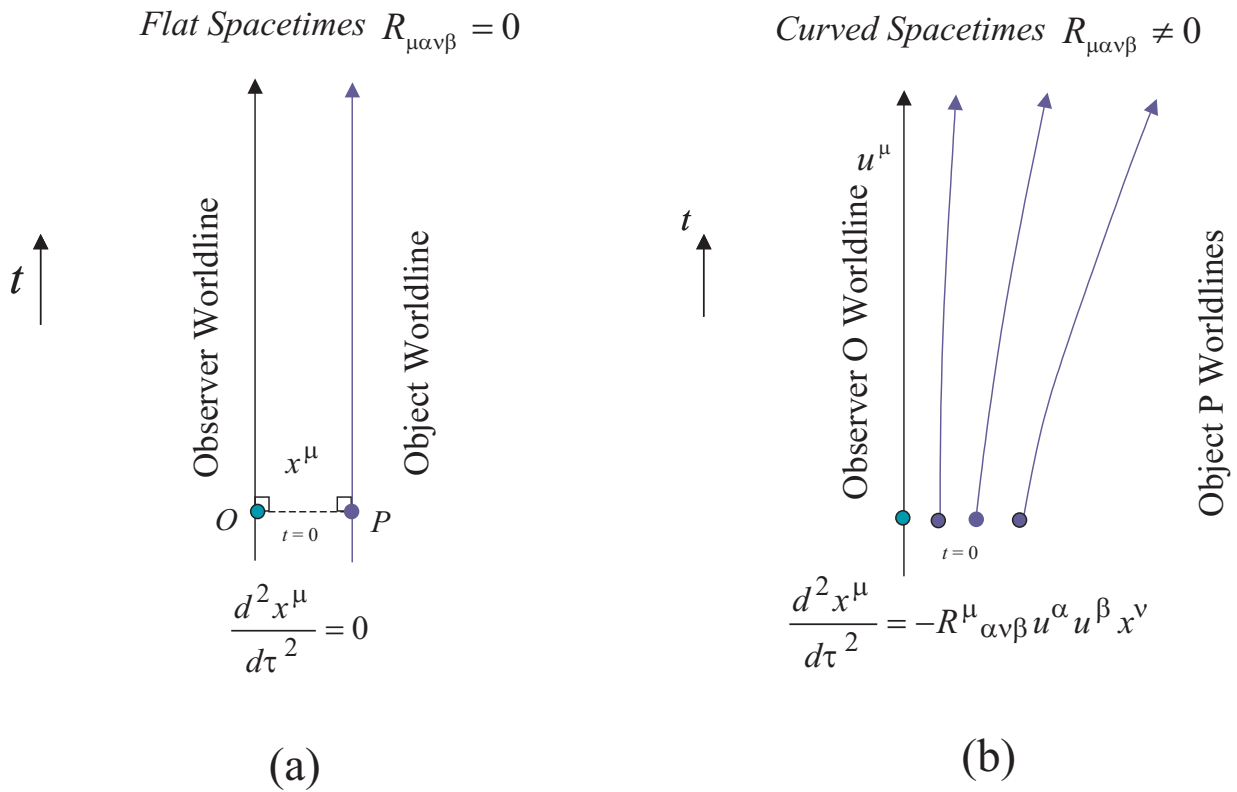}
\end{center}
\caption{Euclid's parallel-line axiom applied to the flat spacetime shown
  in (a) implies that the worldlines of two small objects O and P at
  rest with respect to each other at $t=0$, remain parallel to each
  other and never meet. In the curved spacetimes shown in (b),
  however, the two worldlines can either converge or diverge, or, as
  in the case of GWs, can oscillate. The acceleration of an object
  seen by the observer O close to P is proportional to the local
  Riemann curvature tensor, and is also proportional to the
  \textit{separation} $x^\mu$ between O and P, as is the case in all
  such \textit{tidal} effects.}
\label{parallel-world-lines}
\end{figure}

Einstein generalized Riemann's notion of curved $space$ to curved
\textit{spacetime} by the generalization of the curvature of space to
that of the curvature of spacetime, so that
\begin{equation}
R_{ijk}^{l}\rightarrow R_{\mu\nu\kappa}^{\lambda}
\end{equation}
where the Greek indices $\lambda,\mu,\nu,\kappa$ now run from 0
(representing time), to 1, 2, and 3 (representing space). He then
applied the geometrical concepts of Gauss and Riemann to describe gravity:
Matter acts on geometry by curving spacetime, and geometry acts on
matter by determining the paths of test particles \cite{MTW}.
Flat spacetimes still obey Euclid's parallel-lines axiom, now
generalized to the case of two straight and parallel
\textit{worldlines}, i.e., trajectories of noninteracting particles in
space and time, for two small, nearby free objects at rest with
respect to each other, e.g., in outer space, located,  at time $t=0$,
in their proper frames at the two points O and P, as illustrated in
figure \ref{parallel-world-lines}(a). These two straight and parallel
worldlines are also implied by the rectilinear, uniform motion of free
objects obeying Newton's first law of motion.

In this space-time diagram, time, represent by the vertical axis, ascends
vertically, and space is represented in one dimension by the horizontal axis.
Thus one of the objects, say the observer O, is represented by a vertically
ascending, straight worldline, and a nearby object P is represented by the
nearby, parallel worldline. If spacetime were strictly flat (as shown
in figure $\ref{parallel-world-lines}$(a)), these two
straight and parallel worldlines would never meet. However, like the great
circles on a sphere, or geodesics on a pseudosphere, in
curved spacetimes these lines can either converge, or diverge from each
other, respectively, depending on the sign of the components of the
Riemann curvature tensor. An observer fixed on O would see the object
P either come towards him, or go away from him, and would interpret this
motion as being due to an effective `tidal' force acting on P (see
Chapter 1 of \cite{MTW} for a description of this). In figure
$\ref{parallel-world-lines}$(b) we have drawn the specific case where
components of the local Riemann curvature tensor are negative, and the
worldlines diverge from one another. Since both the observer and the
object are simply following their own worldlines\textemdash which are
geodesics in the absence of all other forces on both the observer and
the object\textemdash the equivalence principle still holds, and
consequently this force is proportional to the mass of the object. The
resulting motion of P as observed by O is therefore
\textit{independent} of the mass, composition, or thermodynamic
state of the test object at P, and herein lies the
\textit{geometrical} meaning of the equivalence principle. Moreover,
when the local Riemann curvature tensor between O and P can be
approximated as a constant, this force, like the action of the Moon's
gravity on the tides, \textit{increases linearly} with the distance
between O and P (see figure $\ref{parallel-world-lines}$(b)), and
indeed, the use of the name `tidal' for this effective force comes
from the tidal force on the oceans by the Moon's gravitational field.

It is important to note that while the worldline illustrations in
figure $\ref{parallel-world-lines}$ are a visualization of
geometry, and its connection to gravity and gravitational tidal
forces, they are only a convenience, and do not represent any
frame that can be achieved physically. As drawn, the perspectives
are that of an Observer who is removed from the spacetime, and
standing outside of it looking in. This is unphysical. Every
experimental measurement is made through an experimental
apparatus\textemdash which may be as simple a device as a pair
eyes used to see the motion of an object P a small distance
away\textemdash that is, by necessity, a physical object. As such,
the apparatus \textit{must} lie along a worldline in the spacetime
of the universe, and cannot be removed from it. \textit{No}
experimental measurements can be done by a fictitious observer
`suspended' outside of the universe.

Of particular interest to this paper is the case when a
`gravitational wave' (GW for short) is present in the spacetime. In
this case the worldlines shown in figure
$\ref{parallel-world-lines}$(b) will periodically converge and diverge
from each other, due to the fact that the curvature of spacetime is
alternating between positive and negative values as the GW passes over
the observer O and object P. Such distance changes between the two
objects can be measured by sending a light beam out from O, and
reflecting it by means of a mirror at P back to O. The distance
between the two objects, measured in this way, will also change
periodically in time when the GW passes over them. Once again the
observer O would interpret the motion of the object P as due to an
effective `tidal' force which, in the limit of long-wavelength GWs,
also grows linearly with the distance separating O and P (see Chapter
35 of \cite{MTW} for a more detailed description).

Because of the experiment by Pound and Rebka \cite{PR} to measure
the gravitational redshift, along with the experiment by Collela,
Overhauser and Werner (COW) \cite{COW} to demonstrate through
neutron interferometry that the gravitational potential of the Earth
induces a quantum phase shift, the general relativistic effect
most familiar to the Atomic, Molecular, and Optical (AMO)
community is the gravitational redshift; GWs and their interaction
with matter are not nearly as well known. There is thus an
unfortunate tendency among the community to try to understand all
general relativistic effects, including those due to GWs, in terms
of the redshift. This cannot be done. If the redshift were used in
an attempt to understand the physics of GWs, it would be like
trying to understand electromagnetic (EM) waves using the scalar
potential only; most of the essential physics would be lost.

Because of the AMO community's unfamiliarity, we present here
a review of GWs. This review is necessarily brief; the
literature on GWs, their generation, propagation and interaction with
matter, is vast (see \cite{MTW} and \cite{Thorne}), and a detailed
presentation on the topic cannot be done within this paper. We will
focus here on the physics that underlie GWs, and will delay the
presentation of a mathematical review to \textbf{Appendix A}.

The existence of GWs is one of the most fundamental predictions of
general relativity. Unlike the Newtonian gravitational potential,
which can be obtained from general relativity by taking the
appropriate non-relativistic and weak gravity limits, GWs do not have
an analog in Newtonian gravity. This is because like
electrostatics, Newtonian gravity is a static theory that cannot
encompass wave propagation. Indeed, the analysis of the
gravitational redshift, as measured in the Pound-Rebka experiment,
could be done using only special relativity and the equivalence
principle; it does not need the full structure of general
relativity. Just as it took the discovery of Amp\`ere's
and Faraday's Laws to lead Maxwell to predict the existence of EM
waves, it took Einstein's generalization of Newtonian gravity to
predict the existence of GWs, and to describe their properties.

\begin{figure}[ptb]
\begin{center}
\includegraphics[width=0.75\textwidth]
{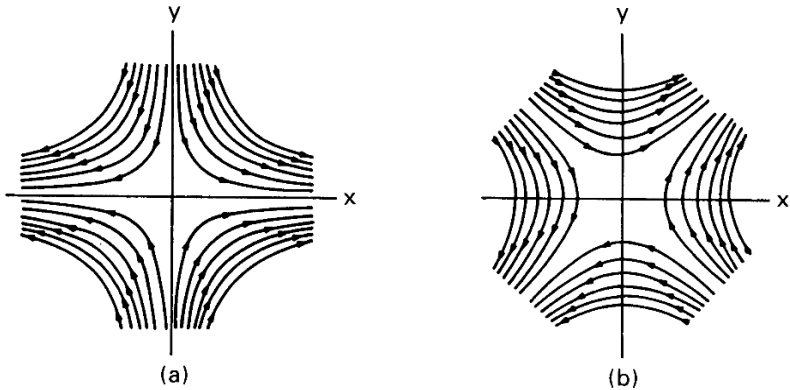}
\end{center}
\caption{The force-field lines for the $+$ polarization (a) and $\times$
  polarization (b) of a GW (from \cite{Blair}). These patterns are the
  snapshots of anisotropies of space, or `strains', induced by
  the GW in the spacetime manifold, taken at $t=0$ by an observer
  located at the centre, i.e., snapshots of the patterns of motion of
  an ensemble of freely-falling, noninteracting test particles
  distributed uniformly over the $(x,y)$ plane. These time-varying
  patterns of anisotropies cause time-varying phase shifts to appear
  both for light waves in Michelson-type interferometers, such as in
  LIGO, and also for matter waves in Mach-Zehnder-type
  interferometers, such as in MIGO.
}
\label{GW-Polarizations}
\end{figure}

Many, but not all, of the underlying properties of GWs in the linear
approximation can be understood  \cite{Comment-1} in analogy with EM
waves (see \textbf{Appendix A}). Like EM waves, GWs have two physical
polarizations\textemdash the $+$ polarization and the $\times$
polarization\textemdash but unlike an EM wave, which is a spin 1 field and
can be represented by a vector potential $A_\mu$, a GW is instead a
spin 2 field and must be represented by a second-rank \textit{tensor}
$h_{\mu\nu}$. This tensor is analogous to a local \textit{strain}
induced on spacetime by a GW, and is a measure of the size of the
`ripples' in spacetime caused by the passage of the GW. Like $A_\mu$,
$h_{\mu\nu}$ is `gauge' field, with the choice gauge being a choice of
\textit{coordinate} systems, rather than a choice in a $U(1)$ phase
factor. The gauge invariant object corresponding to the field strength
$F_{\mu\nu}$ for EM is the Riemann curvature tensor
$R^\lambda_{\mu\nu\kappa}$, constructed from second derivatives of
$h_{\mu\nu}$. While it is possible to choose a gauge where $h_{00} \ne
0$\textemdash the term that would generate gravitational-redshift-like
effects\textemdash this would not be the minimal gauge for the
GW. Making this gauge choice would be analogous in EM to making a
gauge choice where $A_0\ne0$ instead of the usual \textit{radiation}
gauge: $A_0 = 0$ and $\vec\nabla\cdot \vec A = 0$. Moreover, since the
potential $h_{00}$ can describe at most one of the two physical
degrees of freedom of the GW, attempting to understand the effects of
GWs on matter using this $h_{00}$ through a gravitational redshift
argument would miss much of the underlying physics. Both degrees of
freedom for the GW must be taken into account, and this is most easily
done by choosing a minimal gauge. This gauge\textemdash which is
analogous to the radiation gauge for EM waves\textemdash is the
transverse-traceless (TT) gauge: $h_\mu^\mu =0$, $\nabla^\mu
h_{\mu\nu} =0$, and for GWs propagating on flat (`Minkowski')
spacetime, $h_{0\mu}=0$. In particular, the gravitational potential
vanishes since $h_{00}=0$. There can be no gravitational-redshift-like
effects; only the spatial `strain' components $h_{ij}$ are
non-vanishing. Like $A_\mu$, in this gauge $h_{ij}$ can be represented
as plane waves with two polarization vectors $\epsilon^{(+)}_{ij}$ and
$\epsilon^{(\times)}_{ij}$, shown graphically on figure
$\ref{GW-Polarizations}$.

One aspect of GWs that \textit{cannot} be understood in analogy with
EM waves is the interaction of GWs with matter. The reason for this
goes beyond the obvious differences in the tensorial ranks of
$h_{\mu\nu}$ and $A_\mu$. Unlike EM waves, GWs, like all gravitational
effects, \textit{cannot} be screened. As mentioned above, \textit{all}
physical observers must be part of spacetime, and cannot be removed from
it. When a GW passes through a physical system, it acts on
\textit{all} parts of the system, including the observer, and
consequently, only \textit{differences} in the motion between the
observer (O in figure $\ref{parallel-world-lines}$) and the observed
(P in figure $\ref{parallel-world-lines}$) can be measured; the
`absolute' motion of either one cannot (see discussion in
Chapter 35 of \cite{MTW}).

If only \textit{relative} motion can be measured when a GW is
present, what coordinate system should be chosen? A description in
words of this coordinate system has been given by Thorne
\cite{Thorne}, and has been outlined by Synge \cite{Synge}, and de
Felice and Clark \cite{deF} for general spacetimes. More recently,
Speliotopoulos and Chiao \cite{GLF} has explicitly constructed
this coordinate system for general spacetimes in the limit of
linearized gravity. Its construction is based on the following
physical constraints \cite{Comment0}: Every physical particle
travels along a worldline in the spacetime of the universe. Every
measurement of the physical properties of the test particle by an
observer must be done using an experimental apparatus. Before the
observer can take measurements with this apparatus, he must first
choose a local orthonormal coordinate  system. In curved
spacetimes, this involves the construction of a \textit{local}
orthonormal coordinate system (a tetrad  frame) \cite{Synge}.
Naturally, this coordinate system will be fixed, say, to the
centre of mass of his experimental apparatus, and will thus
propagate in time along the worldline of the apparatus as well.
The observer uses the coordinate time of the physical apparatus to
measure time, which, because he or she will not be moving relative
to the apparatus, is also his or her proper time. Thus the time
axis of the coordinate system he or she has chosen will always be
tangential to his or her own worldline. The position\textemdash
which can be of finite extent\textemdash of the test particle is
measured with respect to an origin fixed on the apparatus, and is
the shortest distance between this origin and the particle.
However, because the apparatus travels along its worldline, the
origin of the coordinate system will also travel along a worldline
in the spacetime. When the rate of change of the
position of the particle is measured at two successive times, the
\textit{relative} four-velocity of the particle with respect to
the apparatus will naturally be obtained (see \cite{GLF} for this
construction).

As described, it would seem that the observer simply constructs his or
her usual laboratory frame, and it is difficult to see the novelty of
this construction. The construction is done, however, in a \textit{curved}
spacetime, and as a result, in this coordinate system the observer O
sees the object P in figure $\ref{parallel-world-lines}$(b) undergoing an
acceleration dependent upon the local Riemann curvature tensor of the
spacetime. When no other forces are acting on the object, it is
straightforward to show that when P is close to O, this
acceleration is
\begin{equation}
\frac{d^{2}x_{i}}{dt^{2}}=-R_{0i0j}
x^j=\frac{1}{2}\frac{d^{2}h_{ij}}{dt^{2}}x^{j},
\label{geodesicdeviation}
\end{equation}
where $R_{0i0j}$ are components of the Riemann curvature tensor.
The second equality holds for the special case when long-wavelength
GWs (see \textbf{Appendix A})\textemdash which causes the deviations
$h_{ij}$ of the metric from flat spacetime\textemdash are
present. This equation is obtained by taking the usual
\textit{geodesic} equation for the object P, subtracting from it the
geodesic equation for the observer O, and expanding the result to
first order in $x^i$, the distance separating O and P shown in figure
$\ref{parallel-world-lines}$(b). Equation $(\ref{geodesicdeviation})$
is the geodesic \textit{deviation} equation of motion, and describes
the \textit{difference} in the motion between two nearby
geodesics. Note that the acceleration in
Eq.~$(\ref{geodesicdeviation})$ is proportional to 
$R_{0i0j}$, \textit{and is a gauge-invariant quantity}. Consequently,
although the explicit dependence of this acceleration on $h_{ij}$
depends on the choice of the TT gauge for the GW, the
\textit{dynamics} (or acceleration) of the object described by
Eq.~$(\ref{geodesicdeviation})$ will not, in the end, depend on this
gauge (or coordinate) choice.

Turning now to geometry and quantum systems, Berry pointed out that
non-Euclidean geometry can often be encountered in quantum mechanics
\cite{Berry} as well as classical mechanics. For example, a spin-half
particle can be constrained by a strong magnetic field to always
remain parallel to the direction of the field. When the direction of
the magnet that produces this field is slowly rotated through space
around a closed conical circuit, there results a rotation of the
spin through the angle $\phi$ after the completion of a circuit. After
this round trip, the spin picks up a quantum phase factor which is a
consequence of the non-Euclidean geometry of the Bloch sphere (see figure
$\ref{parallel-world-lines}$(b)).

This phase factor can be explicitly seen through the Gauss-Bonnet theorem,
which states that $\phi=\Omega$, where $\Omega$ is the solid angle
subtended by the circuit with respect to the centre of
sphere. There therefore results a quantum phase shift $\Delta\Phi$
experienced by the $rotated$ spin after this round trip, where
$\Delta\Phi=m_{s}\phi=m_{s}\Omega$ for the case where the spin stays
adiabatically in its eigenstate $\left|m_{s}\right\rangle $. After
the circuit is completed, the spin thus picks up a phase factor of 
$\exp\left(im_{s}\phi\right)=\exp\left(im_{s}\Omega\right)$ that can
be measured in quantum interferometry. This is Berry's `geometric
phase factor', or `anholonomy', which is similar to the Aharonov-Bohm
phase factor picked up by an electron after it traverses a circuit on
the surface of a sphere surrounding a Dirac monopole.

Irrespective of the great successes of differential geometric concepts
in classical physics, since all objects at a fundamental level obey
quantum mechanics, the notion of `worldlines', or of `geodesics', or
of `trajectories' in general, is problematic because of the
uncertainty principle. As Bohr has taught us, we must fundamentally
abandon the notion of \textit{classical trajectories} in quantum
mechanics. Nevertheless, from Berry's work it is well known that
geometric concepts are fruitful in quantum mechanics as well. The
crucial question then arises: How does one \textit{operationally measure}
spacetime curvature in a quantum world? As a corollary, how does one
know whether a spacetime is \textit{precisely} flat or not? The
answer: By means of the interference of \textit{quantum} test
particles using matter-wave interferometry, as these test particles
move through spacetime. These test particles are \textit{quantum} in
the sense that in the interferometer, we cannot know, even in
principle, \textit{which path}, in the quasi-classical limit, a given
particle actually took. Hence we are strongly motivated to examine the
problem of matter-wave interference as a means of quantifying the
problem of the \textit{quantum measurement} of the curvature of
time-varying spacetime geometries, and in particular, the specific case
of those associated with gravitational radiation. This is closely
related to the more general question: How does $gravitational$
radiation interact with $quantum$ matter? As we shall see below, this
question already arises in an interesting way at the level of
$nonrelativistic$ quantum matter interacting with \textit{weak,
  linearized} GWs. This problem is a $linear$ one both in the matter
and the field sectors of the theory, and should therefore be
calculable using linear response theory.

\begin{figure}[ptb]
\begin{center}
\includegraphics[width=0.5\textwidth]{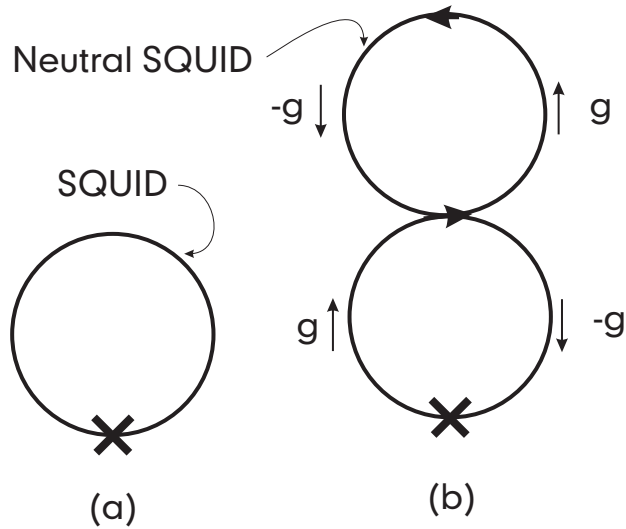}
\end{center}
\caption{SQUIDs as detectors of long-range fields. `X's denote Josephson
junctions. (a) A superconducting SQUID is a sensitive detector of magnetic
fields through the Aharonov-Bohm effect. (b) A neutral SQUID (e.g., using
superfluid helium) could also in principle detect gravitational radiation
through a gravitational version of the Aharonov-Bohm effect. As a GW
with $+$ polarization passes over the interferometer propagating from
left to right, the left and right sides of the SQUID, which are spaced
by half a wavelength of a GW, causes tidal forces that reverse in sign
from left to right. These forces induce a time-varying phase shift,
which can be detected by means of the Josephson junction.}
\label{SQUID}
\end{figure}

One of the authors (RYC) started thinking about this problem two
decades ago \cite{Chiao1982}, when he asked the question: If the
superconducting quantum interference device (SQUID), an early form of
a matter-wave interferometer, can very sensitively measure the
Aharonov-Bohm phase, what would happen if one were to replace the
\textit{charged} particles inside the SQUID by \textit{neutral}
particles? Is there some magnetic-like field
leading to an analog of the Aharonov-Bohm effect, so that the
neutral version of the SQUID can also sensitively measure this
field? At that time, the answer seemed to be that the Lense-Thirring
field of general relativity was this magnetic-like field. A natural
extension of this question was: If one could sensitively measure
\textit{stationary} gravitational fields by means of quantum
interference, could one also sensitively measure
\textit{time-varying} gravitational fields, i.e., gravitational
radiation? One answer: The quadrupole (or tidal) symmetry of
gravitational radiation would require that the neutral SQUID be
twisted into a figure-8 shape before it could detect such radiation
(see figure \ref{SQUID}(b)).

Since the neutral SQUID, viewed as a macroscopic quantum system, would
become, for weak gravitational fields, \textit{linear and reciprocal}
in its response to GWs, the question naturally arose: Could this
figure-8 antenna also \textit{generate}, as well as \textit{detect},
GWs? Recently, RYC suggested a simplification of these earlier ideas,
which eliminated the hard-to-make Josephson junctions
\cite{Chiao}. Quantum systems such as the quantum Hall fluid, a
\textit{ferromagnetic} material that responds linearly and
reciprocally to external perturbations, could
lead to a kind of `quantum transducer action', in which conversion of
GWs into EM waves, \textit{and vice versa}, could in
principle occur. This coupling originates from an extension of the
usual minimal coupling rule to include the coupling of spin to curved
spacetime \cite{Chiao}.

For example, the spin of the electron in a quantum Hall fluid can be
viewed as a handle, by which it can be twisted around periodically in
direction by a GW, thus causing it to pick up a periodic Berry
phase. (The electron spin, which is a gyroscope-like object, locally
undergoes \textit{parallel transport} in the presence of a GW.) The
Berry phase in turn induces macroscopic quantum flows in this charged
quantum fluid which radiate EM waves. This \textit{macroscopic
  quantum} process, in which a GW generates an EM wave, is
\textit{both linear and reciprocal}. Because of time-reversal
symmetry, the time-reversed process in which an EM wave generates a
GW, must also occur with equal power conversion efficiency; this
efficiency can be at most unity \cite{Comment}. This reciprocity
principle suggests the possibility of a Hertz-like experiment for both
generating and detecting high-frequency GWs, in which, for example,
microwave EM radiation is first converted into a GW by one sample of
the quantum fluid, and then the generated microwave GW is
back-converted into EM microwaves by a second sample of the same
quantum fluid. (Faraday cages would prevent the usual EM coupling
between the generation and detection parts of the apparatus.)
A first attempt at this type of Hertz-like experiment was performed
using YBCO, a superconductor instead of a quantum Hall fluid, at
liquid nitrogen temperature as the quantum sample \cite{ChiaoYBCO}. An
upper limit on the power conversion efficiency for the quantum
transducer action of the YBCO sample was placed at 15 parts per
million. YBCO is a zero-spin superconductor, however. Better choices
might have been non-spin-zero materials like the spin-triplet
superconductors, such as Sr$_2$RuO$_4$ \cite{Mackenzie}, or the
ferromagnetic superconductors, such as URhGe \cite{Paulsen}.

\section{Matter-wave interferometry}

Our primary motivation for studying GWs and their interaction with
matter on the quantum level is based on the observation that
matter-wave interferometers can be very sensitive detectors of GWs from
astrophysical sources. The rest of this paper will be focus on
demonstrating this sensitivity, and exploring the suitability of using
matter-wave interferometers to construct MIGO, the
\textit{M}atter-wave \textit{I}nterferometric
\textit{G}ravitational-wave \textit{O}bservatory.

The gain in sensitivity expected in using matter-wave
interferometry instead of laser interferometry in detecting GWs
can be seen from the following argument: Roughly speaking, since
an atom with mass $m$ `weighs' much more than a photon with
frequency $\omega_{\gamma}$, all other things being equal, a
matter-wave-based Sagnac interferometer will be
$mc^{2}/\hbar\omega_{\gamma}\sim10^{10}$ times more sensitive as
an inertial sensor of Earth's rotation than a light-wave-based
interferometer \cite{Scully}. This fact is well known \cite{DIMA},
and the advantage of an atom-based Sagnac interferometer over a
laser gyroscope has recently been shown experimentally \cite{Kas1,
Kas2}. Although other important factors will also play a role, the
inherent advantage of matter-wave interferometers over light-wave
interferometers is expected to be pervasive, and not restricted only
to the Sagnac interferometer.

Matter-wave interferometry is based on the particle-wave duality of quantum
mechanics. This duality states that any massive object, such as an atom, can,
under certain circumstances, behave like a particle, but can, under different
circumstances, behave like a wave with a deBroglie wavelength
$\lambda_{dB}=2\pi\hbar/mv$, where $v$ is the speed of a
nonrelativistic ($v<<c$) object. Like the phase of light, the
\textit{quantum} phase of an atom can be used to construct an
interferometer using atoms, and we shall show that very sensitive
observatories for GWs, which are expected to be many times smaller
than corresponding light-wave-interferometer-based GW observatories
such as LIGO (\textit{L}aser \textit{I}nterferometer
\textit{G}ravitational-wave \textit{O}bservatory) and LISA
(\textit{L}aser \textit{I}nterferometer \textit{S}pace
\textit{A}ntenna), can be constructed with matter-based
interferometers.

We shall argue here that the rapid technological advances in AMO
physics that have occurred since the pioneering
\cite{Pritchard} work on atom interferometry have now made it possible to
construct MIGO. These technologies did not exist in the 1970s and 80s
when LIGO was first conceived and developed (see \cite{Thorne} for a
history of LIGO); the inherent advantage of matter-wave
interferometers over light-wave ones\textemdash suggested by
the factor $mc^{2}/\hbar\omega_{\gamma}\sim10^{10}$\textemdash either
was not widely known or was not appreciated. Yet due in great part to
this factor, MIGO can not only be many times smaller than either LIGO
or LISA, but can also expand the frequency response range of current
GW observatories to presently inaccessible regions, and can
substantially extend the observational reach of GW observatories as
well. It will also be possible to measure with MIGO other general
relativistic effects that are not accessible to LIGO or LISA.

The study of matter-wave interferometry in connection with GWs was
started by Linet and Tourrenc \cite{Tourrenc}. A few years later,
Stodolsky also studied matter-wave interferometry and its use in
measuring general relativistic effects, including the detection of
GWs. Both of these studies were done on the quantum mechanical
level, and used as their underlying basis the \textit{geodesic}
equation of motion, not the geodesic \textit{deviation} equation
of motion, to calculate the phase shift of individual particles in
both stationary spacetimes, and spacetimes with GWs. These works were
complementary to those by Anandan \cite{Anandan1, Anandan2,
Anandan3}, who focused more on matter-wave interferometry in
stationary spacetimes and the Sagnac effect. He also used the
quantum mechanical framework based on the geodesic equation. More
recently, statements of the impracticality of using atom
interferometry to detect GWs have been made by Kasevich and
coworkers \cite{Kas1}. Studies of GW antennas using quantum
condensed matter systems such as superfluids have also been done,
and were started by Anandan and Chiao \cite{AnanChiao}. The
extension of the quantum-mechanical-based analysis to a
quantum-field-theoretic approach for spin-zero particles was done
by Cai and Papini \cite{CaiPapini}. Later, Bord\'e and coworkers,
in a series of papers \cite{Borde1, Borde2, Borde3}, revisited the
question of matter-wave interferometry and the measurement of
gravitational effects\textemdash both stationary and
nonstationary\textemdash using also quantum-field-theoretic
methods, but now with the Dirac equation on curved spacetimes for a
spin-half particle as their starting point. All of these
quantum-field-theoretic approaches nevertheless resulted in a
phase shift that is similar to Linet-Tourrenc's and to
Stodolsky's.

What has not been well appreciated in the previous work in matter-wave
interferometry and its use in detecting GWs is that gravity, unlike
the other forces of nature, cannot be screened. Thus, when a GW passes
through a system, say a matter-wave interferometer, the GW acts on
\textit{all} parts of the system. Consequently, only
\textit{differences} between particles can be measured \cite{MTW,
  GLF}, and in the long-wavelength limit for GWs, the motion of
particles is described by the geodesic \textit{deviation} equation
Eq.~$(\ref{geodesicdeviation})$, and \textit{not} by the
\textit{geodesic} equation on which previous analyses have been based. This
fundamental feature of GWs is well known in the general relativity
community, and is explicitly exploited in the design of both LIGO and
LISA. In this paper, we will follow LIGO and the standard
general-relativistic approach of using the geodesic \textit{deviation}
equation as a starting point in our analysis of matter-wave
interferometry in connection with GWs. As a consequence, we arrive at
an a phase shift for MIGO that is very much different than those
calculated before based on the geodesic equation of motion. A detailed
comparison of the approaches based on the \textit{geodesic} equation,
such as those listed in the above, versus our approach based on the
geodesic \textit{deviation} equation will be presented elsewhere
\cite{Preparation}; we shall only present a general critique of the
geodesic-equation approaches here.

Unlike LIGO, which uses laser interferometry to very accurately
measure slight shifts in the \textit{position} of a classical test
mass\textemdash the end mirrors of the interferometer\textemdash when
a GW passes by, MIGO measures slight shifts in the \textit{velocity}
of atoms caused by the GW. These velocity shifts in turn produces
small changes in the deBroglie wavelength of the atom\textemdash a
quantum test mass\textemdash and thus in its quantum phase. Like
laser-based interferometers, these phase shifts can then be detected
using interferometry, and we shall show that, assuming equal
shot-noise limits,
\begin{equation}
L_{MIGO}\approx\left\{2B\left(\frac{\hbar\omega_{\gamma}}{mc^{2}}\right)
\frac{c}{2\pi fL_{LIGO}}\right\}^{1/2}\>L_{LIGO},
\label{MIGO-LIGO-Comp}
\end{equation}
where $L_{MIGO}$ is the effective size of MIGO, $B=75$ is the number
round trips the light beam takes within the Fabry-Perot interferometer
placed in each arm of LIGO, $\omega_{\gamma}$ is the frequency of
LIGO's lasers, $m$ is the mass of the atom used in MIGO, $f$ is the
frequency of the GW, and $L_{LIGO}$ is the physical length of one of
LIGO's arms. At 125 Hz, where LIGO I, the current LIGO configuration,
is most sensitive, $L_{MIGO}\approx1.5$ m as compared to $L_{LIGO}=4$
km if a $^{133}$Cs atom is used \cite{Comment2}.  Similarly, a
comparison of a space-based MIGO with LISA leads to a MIGO
configuration that is at least ten thousand times smaller than LISA,
if they both have the same shot-noise sensitivity.

It should be emphasized that at a fundamental level MIGO measures the
local Riemann curvature tensor \textit{no matter what its source
  is}. This can most clearly be seen by drawing in \textit{spacetime}
the paths of an atom passing through an interferometer as was done in
\cite{GRG}; the two possible paths of the atom naturally forms a
closed path in spacetime, similar to the ones is used to define the
Riemann curvature tensor (see figure \ref{euclid-plus-gauss}(b)). It is
thus not surprising that like the well-known Berry's phase
\cite{Berry} and the Aharonov-Bohm effect for GWs \cite{GLF}, the
Riemann curvature tensor encircled by the closed path induces a net
quantum phase shift. However, for GWs we shall see that the additional
action of the mirrors in MIGO on the atom produces a larger phase
shift than that caused by the curvature tensor alone. Thus, in addition
to detecting GWs, MIGO, unlike LIGO, will also be able to measure
directly the local Riemann curvature tensor from \textit{stationary}
sources such as the Earth, Moon, and Sun for the first time, and it
may also be possible to measure the Lense-Thirring field of the Earth
as well.

The main objective of this paper is to present the following: The
concept of MIGO along with its underlying theoretical framework;
an outline of the design of two different types of MIGO
configurations; a calculation of the expected sensitivities of
these configurations; an assessment of their potential as
gravitational wave observatories. We shall also argue as to the
feasibility of constructing a MIGO. To this end, using our
analysis of the underlying physics of MIGO, we have estimated the
specifications of an interferometer that would be capable of measuring
GWs, and have outlined the technologies that could be used in
reaching them. What has \textit{not} been done here is a complete
analysis of the systematic errors of MIGO. Since such an analysis
would depend crucially on the precise structure of the
interferometer, it would not be fruitful to do this without at
least a small-scale MIGO in hand. We have, however, estimated the
\textit{fundamental} limitations to MIGO's sensitivity due to
thermal fluctuations of the mirror. Moreover, as we shall see
below, the various technologies needed to construct MIGO have
already been \textit{separately} demonstrated in various atom
diffraction and interferometry experiments since the early 
pioneering work in the 1990s. What has not been done is the
integration of these components into a complete atom
interferometer that has a design and a sensitivity to detect, measure,
and observe GWs. 

It is unfortunate that misconceptions of atom interferometry, on
the one hand, and laser gravitational wave detection schemes, on
the other hand, are prevalent among both the LIGO and the AMO
communities. Among the AMO community, on the one hand, it is often
believed that it is the light beams in the arms of LIGO's
interferometer that is being acted upon by the GW; the mirrors plays
little role. Precisely the opposite is true. In Thorne's `proper
reference frame', it is the action of the GWs on the end mirrors of
the interferometer, which are suspended 
vertically with piano wires, that is being measured; light is used
simply as a means to measure the shifts in the mirror's position. In
addition, the AMO community often tries to understand the
properties of GWs and their interaction with matter through an
analogy with the gravitational redshift. This also is too naive.
As we have described in the Introduction, GWs are a
\textit{dynamical} effect, while the gravitational redshift is a
\textit{static} effect. Attempting to understand the action of GWs
on matter with the gravitational redshift is like trying to
understand electromagnetic waves using only the scalar potential.
Next, the fundamental difference between geodesic motion and
geodesic \textit{deviation} motion is not appreciated. In
stationary spacetimes, such as that surrounding the Earth, a
global coordinate system can be constructed, and the motion of
test particles can be described using the geodesic equation
\cite{GLF}. In non-stationary spacetimes, such as when a GW is
present, such a coordinate system cannot be constructed, and only
the \textit{relative} motion of test particles can be described,
resulting, in the long-wavelength limit, in the geodesic
\textit{deviation} equation of motion.

Among the LIGO community, on the other hand, it is often thought that
the use of atom interferometry to detect GWs simply involves
replacing the light beams in LIGO with atomic beams. The gain in
sensitivity is then only due to the shorter de Broglie wavelength of
the atoms compared with the wavelength of the light currently used
LIGO; it is still the motion of the mirrors that will be
measured. This viewpoint also is too naive. Atoms are not photons, and
their response to GWs is different than that of photons. As we shall
see below, the \textit{nonrelativistic} atoms used in MIGO are moving
so slowly  that it is the effect of the GWs on the speed of the
\textit{atoms} that will be measured; the motion of the mirrors are a
secondary effect.

We hope that the pedagogical nature of this paper
will dispel these misconceptions.

\section{Background and Review of Research}

\subsection{Matter-wave Interferometry}

We begin with a brief review of the physics underlying matter-wave
interferometers. Detailed reviews can be found in \cite{Berman, PCC}.

\subsection{Principles of Matter-wave Interferometry}

Matter-wave interferometry is based on the particle-wave duality principle of
quantum mechanics, which states that every massive quantum object
characteristics of both a particle and a wave. Which of these two
characteristics it exhibits depends on the properties an experimentalist
wishes to measure or exploit. Using this principle, various
interferometers have been constructed with a wide variety of
sources, beam-splitters, mirrors and detectors (see \cite{Berman,
  Scoles}). Following the COW \cite{COW} experiment in the 1970s, in
the 1990s, Pritchard's demonstration of \textit{atom}
interferometry \cite{Pritchard} using sodium atoms was followed
shortly thereafter by the work of Chu and Kasevich \cite{ChuKas}
with cesium atoms, to perform extremely sensitive measurements
of the local acceleration due to Earth's gravity. Subsequently,
the phase shift of atoms caused by the Earth's rotation was
measured not only with traditional Mach-Zehnder-type interferometers
\cite{Kas1,Kas2}, but also with $^{4}$He and $^{3}$He superfluid
Josephson-Anderson junctions \cite{Packard1, Packard2} as well.
More recently, matter-wave interference of fullerene (C$_{60}$)
molecules has also been reported \cite{Zeilinger}.

Because of the particle-wave duality, every particle possesses a
deBroglie wavelength $\lambda_{dB}$, and, due to its wavelike nature,
massive particles can be diffracted, reflected, and coherently
beam-split\textemdash all the properties of light that are needed for
light-wave-based interferometry. However, \textit{unlike} light,
$\lambda_{dB}$ depends explicitly on the \textit{speed} of the
particle, and its wavelength can be altered continuously. Just as
important, a slowly moving massive particle responds to the presence
of gravity much more sensitively than light, as we shall see below.

Like light-wave interferometry, matter-wave interferometers can be
divided into three distinctive parts: the source emitting the
interfering particle, the `atom optics' consisting of
beam-splitters and mirrors, and the detector. In a generic
matter-wave interferometer a source emits particles\textemdash
either in a continuous stream or in bunches\textemdash which is
then split into two (or more) paths at the beam-splitters. (It is
important that the particle beam is split \textit{coherently} so
that it is not possible\textemdash due to the superposition
principle of quantum mechanics\textemdash to determine
\textit{which path} the particle will travel along.) Mirrors are
then used to change the direction of the beams so that they
are eventually recombined and detected. Like the beams of light in
light-wave interferometers, the particles passing through the
interferometer picks up a phase shift
$\Delta\phi=\left(S_{cl}^{\gamma_{1}}-S_{cl}^{\gamma_{2}}\right)/\hbar$,
but now $S_{cl}^{\gamma_{1}}$ and $S_{cl}^{\gamma_{2}}$ are the
action for a particle travelling along the two
\textit{quasi-classical} paths $\gamma_{1}$ and $\gamma_{2}$ of
the two arms of the interferometer.

\subsection{Supersonic Sources}

Supersonic sources were first developed in the 1960s by chemists for
the study of chemical reactions (see Chapter 2 of \cite{Scoles}), and
they have a number of properties superior to the effusive sources
(ovens) that are more often used by physicists. While the velocity
distribution of atom beams from effusive sources is essentially a
Maxwellian one that is fixed by the temperature of the oven, the velocity
spread of supersonic sources are much narrower. Fractional velocity
spreads of 1\%\textemdash0.3\% for helium\textemdash has been achieved
from continuous supersonic sources, and even smaller fractional
velocity spreads can be achieved using pulsed sources. Nearly
monoenergetic beams with very high intensities can thus be
formed. Pritchard, for example, has produced $10^{21}$
atoms/cm$^{2}$/s/sr sodium beams \cite{Pritchard}, and Toennies has
produced very cold helium beams with $1.5\times10^{19}$ to
$1.5\times10^{20}$ atom/sr/s \cite{PrivateToennies}. Throughputs of
atoms, measured in atoms per second, in atom interferometers using
supersonic sources are thus orders of magnitude larger than those
obtainable using either effusive sources, such as those used by
Kasevich \cite{Kas1, Kas2}, or magnetic-optical-trap sources, such as
those used by Chu \cite{ChuKas}. In addition, by seeding the beam with
the appropriate type of atom, the centreline velocity of the atomic
beam from a supersonic source can be increased or decreased. This
flexibility is another advantage that supersonic sources have over
effusive sources.

\subsubsection{Continuous Supersonic Sources and Optical Molasses
  Collimation}

A typical continuous supersonic source functions as follows: A jet of
gas from a high-pressure reservoir escapes supersonically in free
expansion through a nozzle, consisting of a small orifice
typically 10 to 100 microns in diameter, into a differentially pumped
low-pressure chamber that has a larger orifice at its output end
called the `skimmer'. This skimmer has the appropriate geometry so
that it can skim away the hotter outer components of the rapidly
expanding gas jet, thus leaving only the intense, low-temperature,
central component of the atomic beam to enter into another
differentially pumped chamber. Importantly, after expansion, the
collision times of atoms in the beam reduce dramatically, and
effectively they no longer collide with one another. The beam is often
further collimated by a slit at its output end, and it could be collimated
yet again using a second slit at the output end of yet another
differentially pumped chamber before it enters the main vacuum chamber
containing the atom-based device, such as an interferometer (see for
example \cite{Pritchard}). With successive stages of differential
pumping, by means of either diffusion or turbomolecular pumps, one can
maintain an ultra-high vacuum in the main chamber that is often
needed. In addition, using optical molasses techniques \cite{Zeeman},
beam collimation can be achieved without throwing away most of the
atomic beam, as is currently done when collimating slits are used. The
combination of 2D optical molasses collimation with effusive sources
has already been achieved by \cite{Kas1,Kas2}. Its combination with
supersonic sources, as we propose here, can yield the high-brightness
atomic beams necessary for a good signal-to-noise ratio in MIGO.

\subsubsection{Pulsed Supersonic Sources and Optical Molasses Collimation}

Most pulsed supersonic sources function in much of the same way as
continuous supersonic sources, but with the addition of a fast-acting
valve to pulse the beam. Duty cycles of roughly 10$^{-4}$ can be
achieved, and as a result, very high fluences can be obtained. Because
of the short duty cycles, smaller vacuum pumps than those used in continuous
supersonic sources can be used instead. Of particular interest is the
pulsed source developed by Powers et.~al.~\cite{Powers}, which uses
laser ablation of metallic sources such as lithium (see \cite{LAblate}
for a review of laser ablation) to generate the atomic beam. The
high-pressure gas chamber is replaced by the laser plus metal
assembly. As before, optical molasses can be added to reduce the
velocity spread, and thereby collimate the beam.

\subsection{LIGO and the Detection of GWs}

With the advantage of being a scalable design, the great majority of the
current experimental searches for GWs are based on laser interferometers. A
number of research groups located throughout the world \cite{Robertson} are
expecting to begin to collect data soon: GEO600, a German-British
collaboration; VIRGO, a French-Italian collaboration; TAMA300, a
Japanese effort; and ACIGA, an Australian effort. LIGO, the US-based,
international collaboration, is currently collecting data, and LISA, a
space-based laser interferometer system, is currently in the initial
planning stage. We shall focus primarily on LIGO in this paper.

LIGO is a set of three interferometers based at two locations\textemdash
Hanford, Washington and Livingston, Louisiana\textemdash separated by 3020 km
\cite{LIGO-Report}. All three instruments are based on Michelson
interferometers with Fabry-Perot arms. The physical length of the each arm of
the main LIGO interferometer is 4 km, and the Fabry-Perot interferometer
increases the optical path of the arm a factor of $2B=150$. In its
current LIGO I configuration, a 8 W Nd:YAG laser is used with
$\lambda_{\gamma}=1.064$ $\mu$m. At the end of each arm is a massive
mirror suspended vertically within a vacuum chamber; the
location of the mirror assembly for the end mirror must be held in
position within $10^{-10}$ to $10^{-13}$ m with respect to the central
beam splitter of the interferometer. As we shall see
in the next section, the passage of a GW through the interferometer shifts the
position of the end mirrors, thereby causing a net phase shift of the
light, which is then measured. The laser interferometer used by LIGO
thus provides the means of accurately measuring the position of the
mirrors over a large distance.

While each end-mirror could be fixed to the mirror assembly, the
end mirrors would, through their connection with the mirror
assemblies, be mounted to the ground below. The response of the
interferometer to the passage of the GW would then depend on the
material properties of both the ground and the frame of the
interferometer across the 4 km spanned by LIGO's arms. Thus, by
fixing the mirrors to the mirror assemblies, one ends up measuring
the response of the \textit{Earth} to the GW, and not of the
mirrors. If this is done, sources of systematic errors\textemdash
such as changes to the material properties in the Earth along
LIGO's arms\textemdash cannot be controlled. Consequently, instead
of fixing the end mirrors to the mirror assembly, they are
suspended on piano wires, and are thus decoupled as much as
possible from sources of uncontrollable systematic errors. Even
so, at frequencies below 125 Hz thermal noise in the piano wires
will begin limiting LIGO's sensitivity, and below 40 Hz, seismic
noise causes a rapid decrease in sensitivity. At frequencies above
125 Hz, shot noise begins limiting the phase sensitivity of the
interferometer.

Construction of LIGO began in 1996, and the main interferometers were
commissioned in 2001 \cite{LIGO-Report}. The first science runs were
started in June of 2002, and these data are currently being analysed. The
installation of LIGO II\textemdash designed to be used for GW
astronomy\textemdash is expected to begin in 2006.

\section{A \textit{M}atter-wave \textit{I}nterferometric
  \textit{G}ravitational-wave \newline \textit{O}bservatory
  (MIGO)}

In this section we outline two possible configurations of MIGO and the
technologies that could be used to construct them. We also derive the
phase-shifts expected for this MIGO, and use this calculation as a guide in
determining the specifications of MIGO based on current technology.

\subsection{Theoretical Basis}

\begin{figure}[ptb]
\begin{center}
  \includegraphics[angle = 270, width=0.88\textwidth]{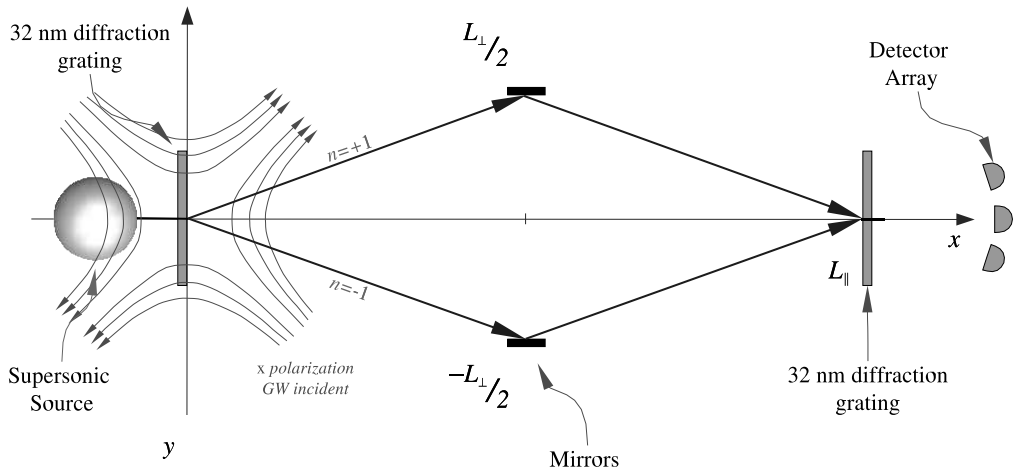}
\end{center}
\caption{Schematic diagram of the horizontal MIGO configuration, with
a GW incident normal to the plane of the interferometer. Only the
  $\times$ polarization contributes to the phase shift; the $+$
  polarization does not. Diffraction orders other than $n=\pm1$ are
  left out for clarity. }
\label{H-V-MIGO}
\end{figure}

\begin{figure}[ptb]
\begin{center}
  \includegraphics[width=0.9\textwidth]{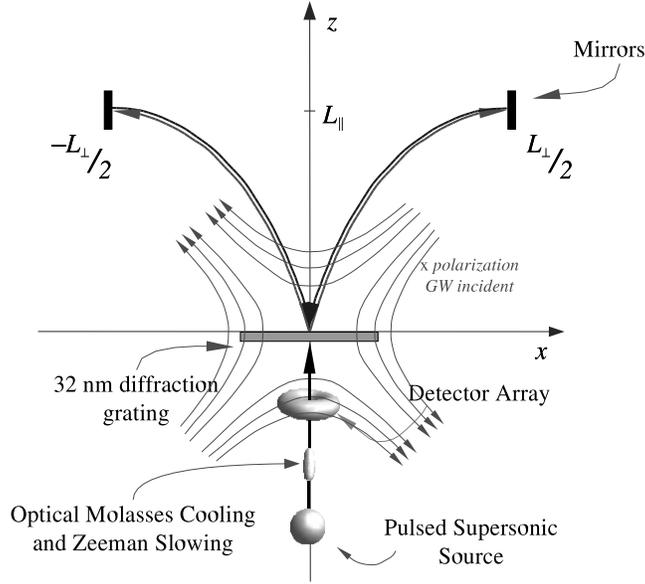}
\end{center}
\caption{Schematic diagram of the vertical MIGO configuration, with a
GW incident normal to the plane of the interferometer. Only the $\times$
  polarization contributes to the phase shift; the $+$ polarization
  does not. Diffraction orders other than $n=\pm1$ are left out for
  clarity.}
\label{g-Dominated}
\end{figure}

In this subsection we outline the physics underlying LIGO and
MIGO, and demonstrate that they are both based on the same underlying
physics, one at the classical level, the other at the quantum
level. The treatment here will be semi-quantitative, and will focus on
the fundamental physics. A complete derivation of MIGO's phase shift
is given in \textbf{Appendix B}, and the reader is referred to
\cite{Thorne} for the derivation of LIGO's phase shift. Like LIGO, we
construct our coordinate system along the lines given in the
Introduction (see \cite{GLF} as well), and we measure the position and
movements of particles with respect to the initial beam splitter (see
figures $\ref{H-V-MIGO}$ and $\ref{g-Dominated}$).

The basic physics underlying LIGO is as follows: Since the mirrors at
the end of each of LIGO's arms are hung as pendula with piano wire,
they are free to oscillate along the interferometer's beam line. As a
GW passes through the interferometer it slightly shifts the position
of both mirrors by a small amount $\Delta x$. This in turn causes a
shift $\Delta\phi_{LIGO}=2\pi B\Delta x/\lambda_{\gamma}$ in the phase
of light (with wavelength $\lambda_{\gamma }$) used in the
interferometer \cite{Thorne, Barish}. (The restorative force of the
pendula on the mirrors can be neglected here.) It is thus the
shifts in the position of the \textit{mirrors} caused by the GW that
is being measured, and not the changes caused by the GW in the
properties of the \textit{light}, as it is often thought
among the AMO community. The light used is only a very accurate
means of measuring shifts in the mirror's position.

As described in the Introduction, the motion of the mirrors in the
interferometer is govern by the geodesic deviation equation
Eq.$(\ref{geodesicdeviation})$. Since the passage of the GW causes
only a small change in the position of the mirror, $\Delta x \approx
L_{LIGO} h_{+}/2$ where $L_{LIGO}= 4$ km is the physical length of an
arm of LIGO, and $h_{+}$ is the amplitude of the $+$ polarization of
the GW; the $\times$ polarization does not affect the motion of the
mirrors appreciably. This shift in the mirror's position changes the
optical path length of the interferometer, and causes a phase shift in the
light used in the interferometer given by $\Delta\phi_{LIGO}= 2\pi B
L_{LIGO} h_{+}/\lambda_{\gamma}$ \cite{Barish, Thorne}. The factor $B$
is the number of round trips that the light beam makes inside the
Fabry-Perot interferometers in LIGO's arms. Strictly speaking this
expression for $\Delta\phi_{LIGO}$ is only applicable for GWs with
frequencies $\sim125$ Hz; at 4 km, the arms of LIGO is equal
to the reduced wavelength of a GW in the upper end of its
frequency response spectrum, and causality has to be taken into
account (see \cite{TysonandGiffard}, page 550). While design
characteristics of LIGO, such as power recycling and shot-noise
modifications by the Fabry Perot arms, will change the sensitivity from
this simple form, it is accurate enough for our purposes in this
section.

If the phase shift for LIGO is proportional to the length of its arms,
how, then, does the phase shift for MIGO depend on its size?
Figures $\ref{H-V-MIGO}$ and $\ref{g-Dominated}$ are drawings of two
possible configurations for MIGO with the force-field lines of a
$\times$ polarization GW drawn in at the origin of our coordinate
system. A complete description of these
configurations will be given in the next subsection. For now, we note
that on a classical level, the motions of the atoms used in MIGO are
governed by Eq.~$(\ref{geodesicdeviation})$ as well. However,
while in LIGO one considers shifts of a mirror's \textit{position}
placed at a distance from the central beam splitter (which is the
origin of LIGO's coordinate system) due to the passage of a GW, in MIGO
one considers shifts in an atom's \textit{velocity} as it travels
through the interferometer due to the GW. Thus, while an atom may have
an initial \textit{velocity} $v_{0}^{i}$ after it passes through the
first beam splitter (in contrast with the end mirror in LIGO which has
an initial \textit{position}), the passage of a GW will cause this
velocity to be shifted by an amount roughly $\Delta v^{i} \sim
v_{0}^{j} \dot{h}_{ij} T/2$ after it has traversed the length of the
interferometer in a time $T$. With a typical $\vert
h_{ij}\vert\sim10^{-20}$, this velocity shift of the atoms is
extremely small, and must be measured on the quantum level by means of
matter-wave interferometry.

Note that the tidal force-field lines sketched in figures
$\ref{H-V-MIGO}$ and $\ref{g-Dominated}$ break the bilateral symmetry
of the interferometers. It is this breaking of the bilateral symmetry
by a $\times$ polarization GW that leads to a non-zero phase shift.

Changes in the atom's speed $|\Delta v|$ result in changes to its
deBroglie wavelength, and hence to cumulative changes in its
quantum phase. Indeed, since the change in the atom's position due to
this speed shift after it passes through the interferometer is
roughly $\Delta L_{MIGO}=2|\Delta v^{i}|T\sim L_{MIGO}T\left|
\dot{h}_{ij}\right|  $ (the factor of $2$ comes from having two arms
in the interferometer), where $L_{MIGO}$ is the effective size of
MIGO, and the transit time $T\sim L_{MIGO}/\vert v_0^j\vert$. Like a
laser interferometer this change in position will cause a phase shift,
$\Delta\phi_{MIGO}=2\pi\Delta L_{MIGO}/\lambda_{dB}\sim
mL_{MIGO}^{2}|\dot {h}_{ij}|/\hbar$, where the \textit{deBroglie}
wavelength of the atom is now used instead of the wavelength of
light. Thus, in contrast to LIGO's phase shift $\Delta\phi_{LIGO}$,
which is proportional to $h_{ij}$, MIGO's phase shift
$\Delta\phi_{MIGO}$ is proportional to the \textit{rate of change}
of the GW amplitude $\dot{h}_{ij}\sim fh_{ij}$.

Although simplistic, this rough derivation of the MIGO phase shift
nevertheless elucidates the underlying physics of MIGO. The above
expression for the MIGO phase shift is close to the exact equation
calculated in \textbf{Appendix B} after a detailed analysis of the MIGO
configurations shown in figures $\ref{H-V-MIGO}$ and
$\ref{g-Dominated}$,
\begin{equation}
|\Delta\phi_{MIGO}(f)|=2\pi\frac{m}{\hbar}Afh_{\times}(f)|F(fT)|,
\label{MIGOPhase0}
\end{equation}
were $A=L_{MIGO}^{2}$ is the effective area of the interferometer, $m$ is the
mass of the atom, $f$ is the frequency of the GW, and $h_{\times}$ is the
amplitude of the $\times$ polarization of a GW. Equation
$(\ref{MIGO-LIGO-Comp})$
follows from Eq.~$(\ref{MIGOPhase0})$, the above expression for $\Delta
\Phi_{LIGO}$, taking $\vert F(fT)\vert \approx1/2$ for freely
suspended mirrors, and setting the shot-noise limits of the phase
shift of MIGO equal to that of LIGO. The resonance function $F(fT)$
depends on the specific configuration of the interferometer, and
measures the resonances between the GW and the interferometer. Precise
forms for $A$ and $F(fT)$ are given in \textbf{Appendix B}. The
corresponding shot-noise-limited sensitivity for MIGO is
\begin{equation}
\tilde{h}(f)_{shot}^{MIGO}=\frac{\hbar}{2\pi
  mAf|F(fT)|\dot{N}^{1/2}},
\label{shot-noise}
\end{equation}
where $\dot{N}$ is the number of atoms passing though the interferometer per
second. In comparison
\begin{equation}
\tilde{h}(f)_{shot}^{LIGO}=\left\{  \frac{\hbar\omega_{\gamma}}{2I_{o}\eta
c}\right\}  ^{1/2}\frac{2\pi f}{\omega_{\gamma}},
\end{equation}
where $I_{o}$ is the power of the laser, and $\eta$ is the photodetector
efficiency (Eq.~123a of \cite{Thorne}). Thus, while $\tilde{h}(f)_{shot}
^{LIGO}$ \textit{decreases} at higher frequencies, $\tilde{h}(f)_{shot}
^{MIGO}$ \textit{increases}. This complementarity between the two
sensitivities is due to a fundamental difference in the signals being
measured by LIGO and MIGO: LIGO measures the \textit{position}, while MIGO
measures the \textit{velocity} of test masses.

Unlike LIGO, MIGO is only sensitive to the $\times$ polarization, not
the $+$ polarization. Also, while the phase shift of LIGO scales with
its \textit{length}, the phase shift of MIGO scales with its
\textit{area}; the larger the area, the smaller the amplitude of the
GW that can be detected. When the transit time is much greater than
the period of the GW, $\vert F\vert \sim 1/2$ for freely 
suspended mirrors; $\vert F\vert \sim \pi fT$ when it is much less
than the period. What is not included in Eq.~$(\ref{MIGOPhase0})$
is the Sagnac effect caused by the rotation of the Earth \cite{Kas1,
  Kas2}, and the phase shift caused by stationary sources of curvature
such as the Earth.  While for the configurations shown in figures
$\ref{H-V-MIGO}$ and  $\ref{g-Dominated}$ these phase shifts are
expected to be very large, they are, however, steady-state, and thus
can be isolated from the time-varying signal caused by a GW.

\subsubsection{Slow Atoms and the Response of MIGO}

With the characteristic speed of an atom from a supersonic source being
in the hundreds of meters per second range, and $T$ in the range of
seconds, we would naively expect MIGO to respond very slowly to the
passage of a GW, which moves through the interferometer at the speed of
light. (However, even for a laser-interferometer-based observatory
such LIGO, which has arms that are roughly equal to the reduced
wavelength of a $10^{4}$ Hz GW, the transit time of light through LIGO
is comparable to the period of the GW as well.) Because of the
cumulative effect of the GW on the phase of the atom, we would expect
$\Delta\phi_{MIGO}$ to be the result of averaging the phase of the
atom over many periods of the GW, and thus to be close to zero. This
does not happen for the following two reasons.

Firstly, half-way through the interferometer the atoms in the beam
hit, and are reflected off, a set of mirrors. These mirrors impart
on the atoms an \textit{instantaneous} force whose strength is
proportional to the \textit{instantaneous} velocity\textemdash
including the changes to the atom's velocity caused by a GW as it
travelled between the beam splitter and mirror\textemdash of the atom
normal to the mirror. From the force-field lines drawn in figures
$\ref{H-V-MIGO}$ and $\ref{g-Dominated}$, the magnitude of this
force will be different for atoms travelling along the $n=+1$ and
$n=-1$ paths, and they will induce a net phase shift between the
atom travelling along the two different paths. This can be seen
explicitly in the derivation of the MIGO phase shift given in the
\textbf{Appendix B}, where the force that the mirrors exert on the
atoms results in a jump condition for the atom's velocity in the
direction perpendicular to the mirror's surface. For
high-frequency GWs this effect of the mirror's impulsive force is
the dominant contribution to $\Delta\phi_{MIGO}$, and is the
reason why $\vert F\vert \sim 1/2$ for freely suspended mirrors
when $T$ is much longer than the period of the GW. At lower
frequencies, when $T$ is much shorter than the period of the
GW, the mirror's impulsive effect is very much reduced.

Secondly, notice from Eq.~$(\ref{geodesicdeviation})$ that due to the
\textit{tidal} nature of the GW, its effect on the motion of the atom
\textit{increases} as the atom moves away from the beam splitter; the
atom `sees' a larger effective acceleration later in its path through
than interferometer than it did at the beginning. Although for low
frequency GWs this results in a relatively small resonance function
$\vert F\vert\sim \pi fT$, the phase shift nonetheless does not average
out to be zero as quickly as one might expect.

The relative sizes of MIGO compared with LIGO given by
Eq.~$(\ref{MIGO-LIGO-Comp})$ would also seem to be counterintuitive. MIGO
makes use of slowly-moving, nonrelativistic atoms to make its
measurements, while LIGO would seem to make use of photons moving at
the speed of light $c$. At first glance it would seem that MIGO should
be \textit{less} sensitive to GWs than LIGO by some power of
$v_{atom}/c$. This, however, would be an erroneous argument. As
outlined in \textbf{3.4}, it is \textit{not} the effect of GW on the
\textit{light} used in the laser interferometer that is being
measured in LIGO; it is the effect of the GW on the \textit{test
  masses} (the \textit{end mirrors}), \textit{which are at rest}, that
is being measured. Indeed, it is precisely because the atoms are
\textit{slowly moving}, i.e., \textit{nonrelativistic}, that the
effect of GWs on their motions are much more readily measurable than
their effect on light. A GW can readily change the speed of
nonrelativistic atoms, but it cannot change the speed $c$ of light.

As a consequence, we would expect the \textit{slower} the atom is, the
more \textit{sensitive} MIGO will be, as long as the atoms in the beam
do not decohere because of long transit times through the
interferometer. As counterintuitive as this conclusion may be, it can
be arrived at through the following thought experiment. Suppose we
replace the slowly moving, nonrelativistic atoms in MIGO by a series of
faster and faster atoms until the atoms in the beam approach the speed
of light. In this ultrarelativistic limit, the atoms behave much like
the light used in LIGO, and it is apparent the horizontal MIGO
configuration would be similar to LIGO, but with mirrors that are
rigidly attached to the frame of the interferometer, instead of the
freely moving mirrors in LIGO. Since it is well known that the
sensitivity of LIGO decreases greatly if
rigidly mounted mirrors are used instead of suspended ones, MIGO's
sensitivity to GWs when ultrarelativistic atoms are used is expected
to be much worse than if slowly moving atoms are used. Seen another way,
because the speed of ultrarelativistic atoms are so close to the speed
of light, changing their speed by even a small amount requires a great
deal of energy. The speed of nonrelativistic atoms, on the other hand, are
much more readily changed, and for the same amount of energy, a GW can
shift the speed of a slow, nonrelativistic atom much more than the
speed near $c$ of an ultrarelativistic one, and thereby can cause a much
larger phase shift for the nonrelativistic atom. To summarize, we
would expect that the sensitivity to detecting GWs to
\textit{decrease} if ultrarelativistic particles\textemdash which
behave like photons\textemdash are used in MIGO instead of slowly
moving, nonrelativistic atoms. However, the slowly moving atoms must
not decohere due to collisions during the transit times through
the interferometer. 

\subsubsection{Comparing MIGO with Other Approaches to the Detection
  of GWs using Matter-wave Interferometers}

Equations $(\ref{MIGO-LIGO-Comp})$, $(\ref{MIGOPhase0})$, and the
conclusion that the sensitivity of atom-based interferometers to GWs
is \textit{larger} in comparison to light-based interferometers, are
also surprising when compared to the results of Linet-Tourrenc
\cite{Tourrenc}, Stodolsky \cite{Stodolsky}, Cai-Papini
\cite{CaiPapini}, and Bord\'e and coworkers \cite{Borde1,
  Borde2}. Complete criticisms of these approaches will be given
elsewhere \cite{Preparation}. We note here simply that the approaches
taken by both Linet-Tourrenc and Stodolsky have their roots in the
\textit{geodesic} equation of motion. Linet-Tourrenc used
the quasi-classical approach based on the relativistic single-particle
Hamiltonian for a relativistic particle derived from the geodesic
equation; Stodolsky based his approach on the relativistic
single-particle action. Both authors do not explicitly define their
coordinate systems. They both find a phase shift of the form
\begin{equation}
\Delta \phi_{geodesic} = \frac{c^2}{\hbar}\int_\gamma h_{ij} p^i p^j
\frac{dt}{E},
\label{S-geodesic}
\end{equation}
from the equation above Eq.~(6.1) of \cite{Stodolsky}, and Eq.~[3.2.3]
  of \cite{Tourrenc}. (To conform with standard units, we have restored the
  requisite factors of $c$ and $\hbar$.) The integral is over a closed
  path $\gamma$  in spacetime that the particle
  takes through the interferometer, $p_i$ is the momentum of the
  particle, and $E$ is its \textit{total} energy including the
  rest mass. In comparison, if we also, for the moment, neglect the
  action of the GW on all parts of the interferometer (which is an
  unphysical assumption), we find the change in the action to be
\begin{equation}
\Delta \phi_{geo deviation} = \frac{m}{\hbar} \int_\gamma
dt\left(\frac{1}{2}\vec{v}^2 -\frac{1}{2}\frac{dh_{ij}}{dt}x^i v^j\right).
\label{S-geo-deviation}
\end{equation}
The difference between $\Delta \phi_{geodesic}$ and $\Delta \phi_{geo
deviation}$ is readily apparent, especially when the nonrelativistic
limit of $\Delta \phi_{geodesic}$ is taken. $\Delta \phi_{geodesic}$
depends on the momentum of the particle and is independent of the
length of the path travelled. Because of this momentum dependence of
$\Delta \phi_{geodesic}$, it is expected that matter-wave
interferometers will become more sensitive the closer to the speed of
light the atom moves; i.e., the more ultrarelativistic the particle
becomes, the larger its phase shift. This can be seen explicitly when the
nonrelativistic limit of Stodosky's expression (Eq.~(6.1)) for the phase
shift of his interferometer is taken, in contrast to our expression
for the phase shift of MIGO. On the other hand, $\Delta \phi_{geo
  deviation}$ increases linearly with the separation $x^i$, as it
should, and the \textit{larger} the interferometer, the larger the
change in the phase, in agreement with the \textit{tidal} nature of
GWs. This linear scaling of the phase shift with the size of the
interferometer is also a result on which LIGO is based. In addition,
unlike previous studies, the effects of the mirrors on the atom's
phase shift was also considered in our analysis. As a
consequence, the phase shift of MIGO calculated in this paper has a
dependence on the velocity of the atoms in the interferometer that is
\textit{opposite} from what Stodolsky calculated, when the
nonrelativistic limit of his Eq.~(6.1) is taken.

It should be emphasized that the differences between $\Delta
  \phi_{geodesic}$ and $\Delta \phi_{geo deviation}$ is due to more
  than the fact that one was derived in a nonrelativistic framework, and
  the other within a relativistic one. This can be seen by comparing
  the equations of motion for the particle used in the two
  approaches. Both Linet-Tourrenc and Stodolsky's approaches use the
  geodesic equation of motion, which, when GWs are present, becomes
\begin{equation}
\frac{d^2 x_i}{dt^2} \approx - \frac{1}{2} \frac{d h_{ij}}{dt}
v^j,
\label{geodesic}
\end{equation}
instead of the geodesic \textit{deviation} equation
Eq.~$(\ref{geodesicdeviation})$. It is, however, well known that GWs
affect \textit{all} parts of the system, and since only
\textit{differences} in positions can be measured, it is the geodesic
\textit{deviation} equation that should be used, and \textit{not} the
geodesic equation. This becomes apparent in Eq.~$(\ref{geodesic})$
with an acceleration for the particle dependent on the
derivative of $h_{ij}$, and thus, from Eq.~$(\ref{Connection})$, on
the connection. Equation $(\ref{geodesic})$ is thus a
\textit{frame-dependent} quantity, and will only have meaning only
when a coordinate system is explicitly chosen. This choice was not
explicitly made either by Linet-Tourrenc or by Stodolsky \cite{Comment3},
nor by Cai-Papini or by Bord\'e and coworkers. This coordinate choice
\textit{is} explicitly made, and is well defined, in the derivation of
the geodesic \textit{deviation} equation of motion,
Eq.~$(\ref{geodesicdeviation})$, and is the reason why the
acceleration of the particle here depends on the Riemann curvature
tensor, a frame-\textit{independent} object. These are the underlying
physical reasons why the above authors obtained the incorrect
expression for the phase shift.

On the quantum-field-theoretic level, Bord\'e and coworkers start with
the Dirac equation in curved spacetime. They used the standard formalism
for quantum field theory in the presence of linearized gravity to
study the phase shift that a GW induces on a generic atom
interferometer using nonrelativistic, spin-half atoms. They
obtained an expression (Eq.~(92) of \cite{Borde1}) for the quantum phase
shift of the atom that has a power-law dependency on $\lambda_{\bot
  dB}$ and $\lambda_{GW}$ that is very much different than ours, and
instead is similar to Linet-Tourrenc's and Stodolsky's. This also led
them to the erroneous conclusion that atom-based interferometers are
at best no more sensitive to GWs than light-based
interferometers. Recall from Eq.~$(\ref{geodesicdeviation})$, that
the \textit{tidal} effects of a GW on an atom \textit{increases} with
the separation distance between the observer and the atom. Bord\'e's
governing Hamiltonian (Eq.~89 of \cite{Borde1}), on which their
analysis was based, does not include these tidal effects of GWs,
resulting in an incorrect expression for the quantum phase shift for
the atom.

\subsection{Description of MIGO Configurations}

The two configurations of MIGO shown in figures $\ref{H-V-MIGO}$ and
$\ref{g-Dominated}$ represent two possible operating extremes. In
the horizontal configuration shown in figure $\ref{H-V-MIGO}$, atoms are
emitted from a continuous supersonic source with a velocity high
enough that the acceleration due to Earth's gravity does not
appreciably alter an stom's path by the time it traverses the
interferometer. Indeed, not shown in the figure is a Zeeman laser
\textit{accelerator}, working oppositely to a Zeeman slower
\cite{Zeeman}, that can accelerate the alkali atoms to even higher
velocities. The atomic beam from the supersonic source is
automatically collimated, and the distribution of velocities of the
atoms in the beam transverse to the beam typically has a $\Delta
v_{t}/v_{s} = 1$\% for most atoms, and can be decreased to $0.3$\% for
helium \cite{Scoles}. Alkali atomic beams from supersonic sources can
then be further collimated by means of 2D optical molasses. The atom
beam then passes through the first beam splitter, a nanofabricated
transmission diffraction grating which, as with the diffraction of
light, splits the beam into different diffraction orders $n$.

As shown in figure $\ref{H-V-MIGO}$, only the $n=\pm1$ orders are
used here. In is important to note that the diffraction grating
splits the atomic beam \textit{coherently}. Consequently, from the
superposition principle of quantum mechanics, \textit{it is not
  possible to determine which path any one atom in the beam will
  take}; there is a finite probability \textit{amplitude} it will take
the $n=+1$ path, and due to the bilateral symmetry of the
interferometer along the horizontal line, the same probability
\textit{amplitude} that it will take the $n=-1$ path. Irrespective of
which path is taken, the most probable trajectory between the initial
beam splitter and the mirror for the atom is determined by
Eq.~$(\ref{geodesicdeviation})$ (see \cite{Berman} for the use of the
Feynman path integral in atom interferometry). The mirror then
reflects the atom by exerting an impulsive force whose strength is
proportional to the velocity of the atom normally incident upon the
mirror. (See \cite{Holst} for recent advances in atom optics, such as
coherent atomic mirrors made from crystals, and the possibility
of using \textit{curved} mirrors to coherently focus and collimate
atomic beams.) Once again the most probable trajectory of the atom
between the mirror and the final beam splitter\textemdash where the
$n=+1$ path and the $n=-1$ path is recombined\textemdash is determined
by Eq.~$(\ref{geodesicdeviation})$; the atoms are then detected using
standard methods \cite{Scoles}. In the absence of gravitational
effects, the bilateral symmetry of the interferometer ensures that the
$n=+1$ path will be the same as the $n=-1$ path, and no phase shift
will be measured. This symmetry is broken when a GW is present,
however; from the force-field lines drawn in figure $\ref{H-V-MIGO}$, we see
that path that the atom takes for the $n=+1$ order will be slightly
different than the one taken by the $n=-1$ order, leading to MIGO's
phase shift.

For the horizontal MIGO configuration, $A=L_{\bot}L_{\|}/2$\textemdash
the actual area enclosed by the interferometer\textemdash in
Eq.~$(\ref{MIGOPhase0})$ (see \textbf{Appendix B}). Since the atoms
travel in approximately straight lines through the interferometer, the
length $L_{\|}$ of this MIGO configuration is determined by its width
$L_{\bot}$, and the ratio of the horizontal velocity $v_{\|}$ of the
atom to its transverse velocity $v_{\bot}$ after the initial beam
splitter (see \textbf{Appendix B}). The sensitivity of a horizontal
MIGO is thus proportional to $L_{\bot}^{2}$.

In the vertical configuration for MIGO shown in figure
$\ref{g-Dominated}$, alkali atoms are emitted from a pulsed
supersonic source, and are then \textit{slowed down} using a Zeeman
slower \cite{Zeeman} to velocities so slow that the acceleration
due to Earth's gravity $g$ now dominates the trajectories of the
atoms through the interferometer, similar to
the atom fountain geometry of Chu and Kasevich \cite{ChuKas}. The
beam is collimated using 2D optical molasses as well \cite{Kas1,
Kas2}. The velocity of the atoms $v_{s}$ along the beam is slowed
to such an extent that after passing through the initial beam
splitter the atoms in the $n=\pm1$ orders now traverse an almost
parabolic trajectory to mirrors placed at the maximum height of
their trajectories. The atoms are once again reflected at the
mirrors, and fall back downward in an almost-parabolic trajectory
where the paths are recombined by the same beam splitter, and
subsequently detected. Once again, the \textit{bilateral} symmetry
of this interferometer along now the vertical line ensures that
the gross motion of the atoms on the left side of the
interferometer is reflected on the right side. The usual parabolic
trajectories of the atoms is shifted slightly by an amount
determined once again by Eq.~$(\ref{geodesicdeviation})$ due to
the passage of a GW, and will again be different for atoms
travelling along the left hand path instead of the right hand
path. It is this \textit{asymmetrical} shift by the GW in the
atoms trajectories that leads to the overall phase shift.

Similar to the horizontal MIGO configuration, $A=L_{\bot}L_{\Vert}$
for the vertical MIGO configuration as well (see \textbf{Appendix B}),
but because the acceleration due to gravity $g$ slows the atom down,
it spends more time in the presence of a GW inside the interferometer
than the atoms in the horizontal MIGO configuration do. Consequently,
$L_{\Vert}=gL_{\bot}^{2}/8v_{\bot}^{2}$, where $v_{\bot}$, the
horizontal velocity of atom after the beam splitter, is inversely
proportional to the periodicity of the diffraction grating used. Thus,
the area $A$ is proportional to the \textit{cube} of the width of
the interferometer, and the sensitivity of the vertical MIGO
configuration is proportional to $L_{\bot}^{3}$, and not to
$L_{\bot}^{2}$ as for the horizontal MIGO configuration, or even to
$L_{\bot}$ as it is for LIGO.

Like LIGO, both MIGO configurations make use of mirrors to redirect
the atomic beams to beam splitters, and like LIGO, the response of MIGO
to GWs will, to a certain extent, depend on how the mirrors are
attached to the frame of the interferometer, and thus how the
interferometer is attached to the Earth. However, because it is
effects of the GW on the interfering particle, i.e., the atoms, that
is being measured in MIGO, and not the mirrors as it is in LIGO,
MIGO is much less dependent on how this attachment is made than LIGO
is. Indeed, from \textbf{Appendix B}, we found that for $fT >>1$,
the response function $1/2\le F(fT)\le 1$, irrespective of how the
mirrors are attached to the interferometer. In fact, the
\textit{lower} limit of $1/2$ that is reached if the mirrors are
freely suspended as in the case of LIGO, and the \textit{upper} limit
of $1$ is reached if the mirror was firmly attached to an
interferometer frame that can be approximated as infinitely
rigid \cite{Comment4}. MIGO thus becomes more \textit{sensitive} the
more firmly the mirrors are attached to the interferometer frame. This
is in direct contrast to LIGO, which become \textit{less}
sensitive. Indeed, if like MIGO we model the connection
of LIGO's mirrors to the interferometer frame as a spring with
resonance frequency $f_0$ and loss factor $Q$, we find from
Eq.~(37.16) of \cite{MTW} that the shift in position of the end
mirrors will be
\begin{equation}
\Delta x = \frac{f^2}{f^2-f_0^2+iff_0/Q}h_{+}L,
\end{equation}
In the limit where $f_0>> f$, when the end mirrors can be approximated
as mounted to an interferometer frame that is infinitely rigid,
$\Delta x\approx 0$, and the LIGO looses all sensitivity to
GWs.

From the sketch in figure $\ref{g-Dominated}$ we see that the vertical
MIGO configuration is a combination of Chu-Kasevich's atomic
fountain interferometer with a Pritchard-type interferometer. Previous
to Chu and Kasevich's work, Zacharias (see \cite{Chu}) attempted to
construct an atomic fountain in the 1950s and failed. Because of the
effusive source used, there was a broad distribution of velocities in
the atomic beam, and as the atoms in the beam climbed up the
gravitational potential of the Earth, they started to slow down, and
other, faster atoms started to overtake them. Scattering with the
faster-moving atoms in the beam took place and coherence was lost. Chu
and Kasevich avoided this by using laser-cooled atomic beams with
narrower velocity distributions, as does MIGO.

\subsection{Potential Sensitivity}

\begin{figure}[ptb]
\begin{center}
\includegraphics[width=0.9\textwidth]{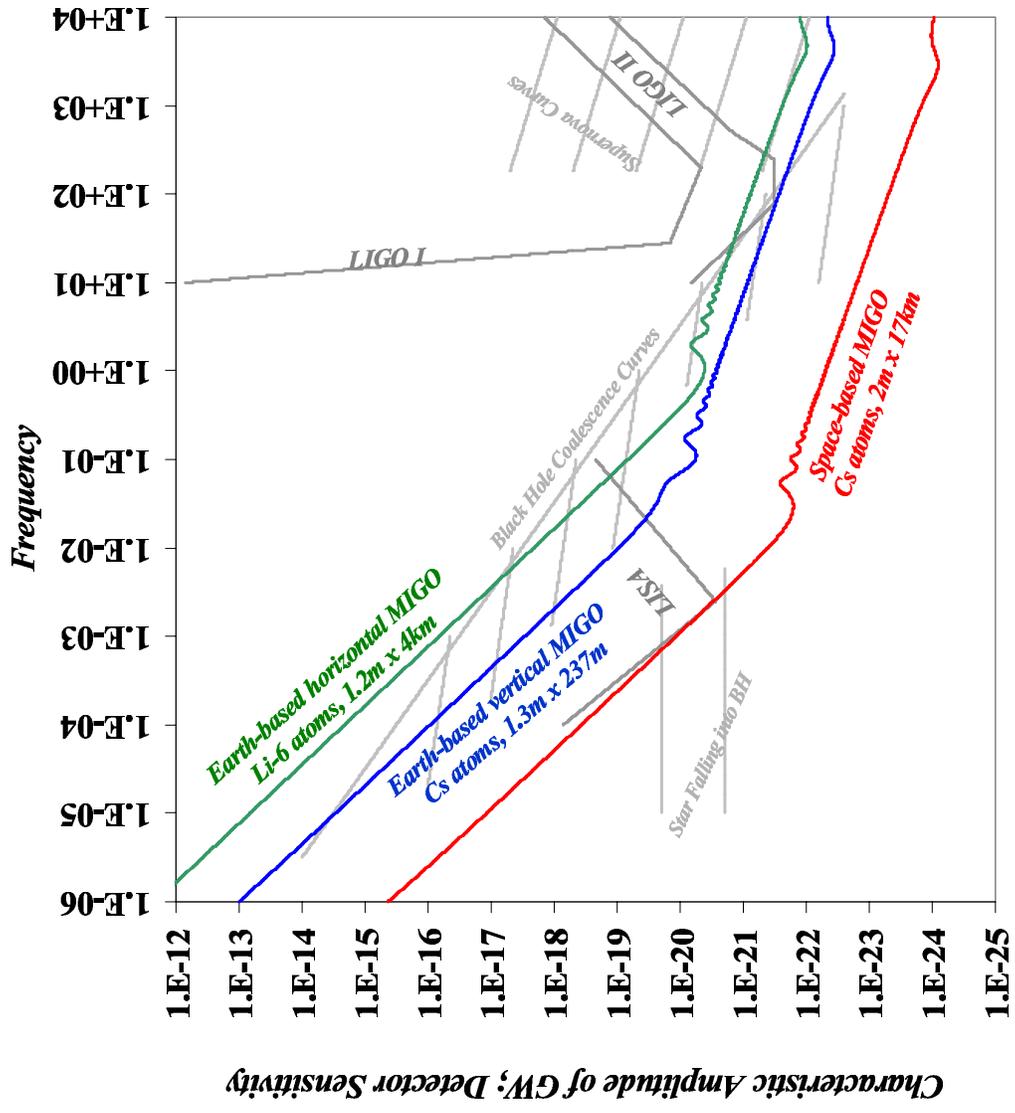}
\end{center}
\caption{The ability of MIGO to detect GWs from various classes of
  burst sources is plotted, and compared with that of LIGO I
  \cite{Thorne}, LIGO II \cite{Fritschel} and LISA \cite{Cutler}. All
  three MIGO configurations are sensitive to a broad range of GW
  frequencies; they thus extend the range of GW observatories into
  frequency ranges not currently accessible.}
\label{h-Burst}
\end{figure}

The effectiveness of both MIGO configurations in detecting GWs can be
seen in figure $\ref{h-Burst}$ where the characteristic amplitude and
frequencies of GWs emitted from burst sources (the strongest emitters
of GWs) such as supernova explosions and the coalescence of black
holes, are plotted along with the amplitude and frequencies of the
\textit{smallest} of these types of signals that are detectable by
MIGO, LIGO (both the current LIGO I configuration and the LIGO II
configuration slated for operation in 2007), and LISA. Two of the
configurations\textemdash the Earth-based vertical and
horizontal\textemdash was designed to have a sensitivity equal to the
best sensitivity of LIGO in its frequency range, and the third
configuration\textemdash the Space-based horizontal\textemdash to have
a sensitivity equal to the best sensitivity of LISA in its frequency
range. The signals from various classes of burst signals were
replotted from Thorne's figure 9.4 \cite{Thorne}; the specifics of
these sources can be found therein. The plots LIGO I, LIGO II and LISA
have been updated from Thorne's original figure, and were generated
from \cite{Thorne}, \cite{Fritschel}, and \cite{Cutler}, respectively,
using Eq.~111 of \cite{Thorne}.

All three graphs for MIGO were generated using
Eq.~$(\ref{shot-noise})$, and a $\dot{N}=10^{18}$ atoms per second (as
compared with $8\times10^{19}$ photons per second for LIGO). While
$\dot{N}=10^{18}$ atoms per second is very large, it is important to
note that what is important is the \textit{number} of atoms per second
through the interferometer, and not the \textit{number
  density}. Indeed, if the number density of atoms is too high, the
atoms in the beam will condense, and beam coherence will be lost. By
suitably constructing the supersonic source, and by 
choosing the proper operating regime for it, it may be possible ensure
that the $10^{18}$ atoms per second needed to achieve figure
$\ref{h-Burst}$ is spread over a large enough cross section of the
beam to prevent condensation. We will comment further on this later in
\textbf{4.4}.

A nanofabricated transmission diffraction grating with a periodicity
of 32 nm was used in calculating figure $\ref{h-Burst}$. Although only
reflection diffraction gratings with this periodicity have been made
by the Nanoelectronic Research Centre at the University of Glasgow
\cite{Glasgow}, by flipping the source and the detector array in
figures $\ref{H-V-MIGO}$ and $\ref{g-Dominated}$, the MIGO
configurations can be modified to use reflection gratings if
transmission gratings cannot be constructed. As with LIGO, because GWs
cannot be shielded, the passage of a GW through MIGO will shift all
its parts, including the mirrors, and the response of the mirrors to
the GW introduces an additional shift in the velocities of the
atoms. However, since the mirrors in MIGOs do not have to be free to
move, they can be constructed so that the effect of the GW on their
motion is critically damped, as they were taken to be when figure
$\ref{h-Burst}$ was graphed.

For the Earth-based vertical MIGO, bosonic $^{133}$Cs atoms from a 2000 K
continuous supersonic source are used in an interferometer with a
width $L_{\bot}= 1.3$ m and a height of $L_{\|} = 237$ m. Zeeman
cooling of the vertical velocity of the atoms from the beam is used to
decrease the velocities of the atom by a factor of 11.6, and
additional 2D optical molasses cooling is used to narrow by a factor
of $10^{4}$ the velocity of the atoms perpendicular to the beam. The
Space-based MIGO, which is depicted as a horizontal configuration,
also uses $^{133}$Cs atoms emitted from a 2000 K continuous supersonic
source, but with 2D optical molasses cooling used only to collimate
the beam. As a result, while $L_{\bot}= 2$ m, the length of the
interferometer is $L_{\|}= 17$ km, as compared to 5 million km for
LISA. (Although a throughput of $10^{18}$ atoms/s is high for a
space-based MIGO, atoms can be recycled from detector to source.) The
final Earth-based horizontal MIGO graph also has a 
$L_{\bot}= 1.2$ m, but uses fermionic $^{6}$Li atoms from a 2000 K
supersonic source in conjunction with a Zeeman laser \textit{accelerator} to
increase the beam-velocity of the atom by a factor of 1.86. By angling
the beam slightly upward, the maximum height of the beam as it travels
across the length $L_{\|} = 4$ km of the interferometer is only $0.41$
m; 2D optical molasses is used to collimate this beam as well. This
MIGO configuration, while not optimal, is designed so that
\textit{whole} interferometer fits within one of the 1.2 m wide
evacuated beam tubes that form the arms of LIGO. Enlarging sections
of the beam tube to $\sim2$ m will allow $^{133}$Cs to be used instead
of $^{6}$Li, resulting in a $37$-fold increase in the Earth-based
horizontal MIGO sensitivity as shown in figure $\ref{h-Burst}$.

In figure $\ref{h-Burst}$ we see that by replacing the lasers in one of
LIGO's arms with the $^{6}$Li horizontal MIGO, the sensitivity of the GW
observatory can be increased by a factor of 10. At $1.2$ m $\times$
$237$ m, the Earth-based vertical MIGO has the same sensitivity as
LIGO II in its most sensitive frequency regime, while at $2$ m
$\times$ $17$ km, the Space-based MIGO has the same sensitivity as
LISA in LISA's operating regime. But all three MIGO configurations are
sensitive to a much wider range of GW frequencies than LIGO and LISA,
and they extend the range of GW observatories into frequency
ranges\textemdash in particular between $0.1$ and 10 Hz\textemdash
not currently accessible by LIGO or LISA. In particular, note the
sensitivity to high-frequency GWs of all three MIGO
configurations graphed in figure $\ref{h-Burst}$, and the gradual
\textit{insensitivity} of LIGO to frequencies above 125 Hz and of LISA
to frequencies above $2\times10^{-2}$ Hz. This is due to the difference
in the signal being measured by matter-wave-based and light-wave-based
interferometers: MIGO measures shifts in the atom's \textit{velocity},
and thus the \textit{rate of change} of the GW amplitude with time;
the larger the velocity shift, the larger the phase shift. This rate
of change \textit{increases} at higher frequencies, and is the
underlying reason the smallest-amplitude GW that three MIGO
configurations can detect in figure $\ref{h-Burst}$ \textit{decreases}
as $1/f^{1/2}$ at high frequencies. In contrast, LIGO (and LISA)
measures the \textit{position} of the end mirrors, which depends only
on the amplitude of the GW, and only indirectly on its
frequency. For the frequency of the GW to be measured, the
smallest-amplitude GW that LIGO can detect must \textit{increase} as
$f^{1/2}$; it actually increases as $f^{3/2}$ because of shot-noise
limits to $\Delta\phi_{LIGO}$. Indeed, if we compare
shot-noise-limited sensitivities,
\begin{equation}
\frac{\tilde{h}(f)_{shot}^{MIGO}}{\tilde{h}(f)_{shot}^{LIGO}}\approx
\frac{\hbar\omega_{\gamma}}{mc^{2}}\left(
\frac{2\dot{N}_{\gamma}\eta}{\dot{N}}\right)  ^{1/2}\left(
\frac{\lambda_{GW}}{2\pi L_{MIGO}}\right)
^{2}|F(fT)|^{-1},
\end{equation}
where $\dot{N}_{\gamma}$ is the rate at which photons enter LIGO.

\subsection{Design Considerations}

The inherent difference in the signal that they
measure\textemdash MIGO measures \textit{velocity} while LIGO
measures \textit{position}\textemdash impacts their constructions,
and the fundamental limits to the size of the phase shifts both
are capable of measuring. Both LIGO and MIGO will be subject to
systematic errors arising from environmental perturbations, such as
thermal and seismic noise; stray electric and magnetic field
gradients; and fluctuations in $g$. Because the temperature of the beam
splitters and the mirrors in both MIGO and LIGO cannot be zero,
temperature-driven changes in their positions and velocities will
induce thermal noise into phase shift measurements. Random seismic
movement of the ground under the interferometer will introduce
vibrational noise as well. However, while thermal noise starts
limiting LIGO's sensitivity below 125 Hz and seismic noise
effectively cuts it off below 40 Hz, their effect on MIGOs
sensitivity is not nearly as great. Thermal and seismic noise have
such a large effect on LIGO because LIGO's end-mirrors must be as
close to a freely-falling mirror as possible.  MIGO, on the other
hand, makes use of beam splitters and mirrors that can be solidly
mounted on an interferometer frame whose material properties can
be well characterized and controlled, because of the relatively
small size of MIGO. Standard vibrational isolation technologies
can thus be used to minimize seismic noise to levels not
attainable by LIGO; for these reasons seismic noise limitations
are not shown in figure $(\ref{h-Burst})$. However,
temperature-driven oscillations on the surface of the
mirrors\textemdash even though they are rigidly mounted\textemdash
will induce random velocity (Doppler) shifts to the atoms, and
subsequently random phase shifts. These oscillations, although not
important for LIGO, will slightly affect the sensitivity of MIGO
at very low frequencies. The effect of these oscillations on MIGO's
sensitivity is estimated in \textbf{Appendix C}, and are included in
figure $\ref{h-Burst}$. Unlike LIGO, the atoms in MIGO are subject to
stray electric and magnetic fields. however, these effects can be made
negligibly small with proper shielding and design. Fluctuations in $g$
are due to density fluctuations caused by seismic waves, and will
limit the low-frequency response of LIGO \cite{Hughes}. Its effect on
MIGO is unknown, and can be eliminated with a space-based system.

At a fundamental level, unlike the photons used by LIGO, the atoms
used in MIGO can interact among themselves; this requires a certain amount of
care to be taken in the design of MIGO. Indeed, fermionic $^{6}$Li was
chosen for the Earth-based horizontal MIGO precisely because as a
fermionic atom it will not scatter with other $^{6}$Li atoms as
readily as $^{7}$Li, a bosonic atom, will. In this sense, cold,
fermionic $^{6}$Li atoms will behave more like the noninteracting
photons in LIGO. Moreover, number squeezed states for fermions, which
are very robust, can greatly reduce the shot noise of the atomic beam
[39], [23]. This correspondingly reduces the required throughput of the
atoms needed for the same sensitivity.

Any collision of the atoms within the beam\textemdash either with other atoms
in the beam, or with stray atoms in the interferometer\textemdash will
introduce random noise in $\Delta\phi_{MIGO}$, and decrease the contrast of
the interference fringes. The cross-sectional area of the beam has to be made
large enough, and the transverse velocity spread of the
beam\textemdash narrowed using 2D optical molasses\textemdash has to
be small enough that this does not occur within the transit time of
the atom through the interferometer. Scattering with the background
gas can be made negligible by using ultra-high-vacuum systems. Next,
the supersonic source must be run at a high enough temperature that
condensation of the gas of atoms into clusters\textemdash either solid
or liquid\textemdash does not occur. At the same time, the pressure of
the supersonic source must be high enough to generate a high intensity
of atoms through the interferometer. Since we are dealing with
atoms which can condense into liquids and solids, the allowed running
temperature and pressure of the source will depend on the
phase diagram of the atom, and the degree of supercooling of the gas
that takes place. These issues are well studied for helium
\cite{Toennies} (although they are not as well studied for the alkali
atoms), and in general there does not seem to be a fundamental
obstacle to finding a set of viable operating parameters. What is more
challenging is the throughput of $10^{18}$ atoms per second used in
generating figure $\ref{h-Burst}$, which is close to the limits of
current supersonic-source technology. However, because of the sensitive
dependence of MIGO sensitivities on its size, a decrease in this
throughput by a factor of 100 can be made up for by
increasing the $L_{\bot}$ of a horizontal MIGO by a factor of
$10^{1/2}=3.16$, or that of a vertical MIGO by a factor of
$10^{1/3}=2.15$; decreases in the atom-beam throughput can thus be
readily compensated for by modestly increasing the size of the
interferometer.

At frequencies below roughly 125 Hz the sensitivity of LIGO
decreases due to thermal noise, and radiation pressure on the end
mirrors \cite{Fritschel}. Photons in the laser impacting the
mirrors not only exert pressure (and thus move) the mirrors, but
because they are not perfect, the mirrors will be heated by the
incident beam. Both factors will cause an overall decrease in
LIGO's sensitivity. One would expect that since $10^{18}$ atoms
per second should pass through MIGO to achieve high sensitivity, the
impact pressure of the atoms on the mirrors, combined with the
heating of the mirrors caused by the inelastic scattering of the
atoms, would limit the sensitivity of MIGO as well. They do not
for the follow reasons.

Quantum interference places stringent limits on the decoherence of the
atom beam as it is diffracted from the beam splitters, and reflected
off the mirrors. (Scattering of atoms in the beam off of background
gas can be shown to be negligible.) In particular, it requires that
there exists no possibility of `which-path' information for the
atom inside the interferometer. This in turn requires that
diffraction of atomic beams off of beam splitters, and their
reflection off of mirrors be elastic and
coherent. \textit{Consequently, an atom can only change the
  centre-of-mass momentum of the beam splitters or mirrors, and cannot
  deposit energy into them.} Since the mirrors and beam splitters of
MIGO can be firmly mounted to the frame, the effects of this
centre-of-mass momentum transfer can be minimized.

A measure of the inelastic versus elastic components of these processes
is based on the Debye-Waller factor $W$ through the intensity ratio
$I/I_{0}=\exp(-2W)$, where $I$ is the diffracted (or reflected)
intensity and $I_{0}$ is the incident intensity. The factor $W$ is a
measure of the fluctuations of the atoms in the crystal that diffracts
the incident atomic beam. For diffracted atomic beams,
$W=\mathcal{B}/a^{2}$, where $a$ is the periodicity of the diffraction
grating, and $\mathcal{B} \sim 0.5$ \AA$^{2}$ at room
temperature \cite{Debye-Waller}. The rule of thumb \cite{Scoles} is that
$W/12<0.1$, for sharp, elastic diffraction patterns to be seen. The
requirement for observing interference is more stringent, however; the
probability of emitting even a single phonon during the diffraction
process must be negligible, and $W/12<0.01$ is required
\cite{Weare}. For nanofabricated diffraction grating and mirrors, $W/12\sim
10^{-7}$ at room temperature and decreases at lower temperatures. Thus
it is highly probable that the \textit{zero-phonon} process is the
dominant one, and therefore quantum phase coherence is expected in the
interferometer.

As stringent as this zero-phonon-emission condition on atom
interferometers may be, atom interferometers of the same type as MIGO
have been successfully constructed before. For example, the
interferometer used by Pritchard, which is very similar to the
horizontal MIGO configurations, uses 1027 K sodium atom beams using
100 nm nanofabricated diffraction gratings. Both the temperature of
the atoms and the size of the diffraction gratings are comparable to
MIGO's. The fact that fringes have already been observed in
Pritchard-type interferometers shows that decoherence due to the
atom-beam-splitter interaction does not prevent interference from
happening.

\subsection{Measuring New General Relativistic Effects}

The underlying advantages of MIGO over LIGO or LISA go beyond being smaller
and more sensitive. It is also possible to explore with MIGO
general-relativistic effects that are not accessible with LIGO.

\subsubsection{Measurements of Stationary Riemann Curvature}

As we have outlined in the Introduction, and have shown in figure
$\ref{euclid-plus-gauss}$(b), the parallel transport of a four-vector
around a closed path is a measure of the local Riemann curvature
tensor of the spacetime. Referring to figure $\ref{H-V-MIGO}$, and
extending the $n=\pm1$ paths that an atom can take into spacetime, it
becomes clear that in MIGO the atom does \textit{precisely} what is
indicated in figure $\ref{euclid-plus-gauss}$(b): It is parallel transported
around a closed path in spacetime, and as a result its phase is
shifted by the local curvature of the spacetime \cite{GRG} through an
anholonomy. Although the delta-function impulsive forces in the
mirrors impart an additional acceleration- and time-dependent phase
shift to the atoms, the Riemann-curvature-tensor contribution still
remains, and will be dominant for low-frequency GWs, and for
stationary Riemann curvature tensors in general. It is thus not
surprising that MIGO is sensitive to the \textit{total} local Riemann
curvature tensor of spacetime, independent of its source, and not just
the fluctuations caused by the passage of a GW, as LIGO is. Indeed,
while in our analysis in \textbf{Appendix B} we have for clarity taken
the acceleration due to gravity to be a constant\textemdash thus
neglecting the contribution of the curvature from stationary sources
such as the Earth\textemdash if we repeat the analysis focusing now
for stationary gravitational fields we find that the horizontal MIGO
configuration will measure a phase shift \cite{GRG}
\begin{equation}
\vert \Delta\phi_{MIGO}^{stat}\vert =
\frac{1}{2}\frac{m}{\hbar}L_{\bot}L_{\Vert}
TR_{0x0y}^{stat},
\label{StatMIGO}
\end{equation}
where $R_{0x0y}^{stat}$ is the \textit{total} local Riemann curvature tensor
from stationary sources (once again the Sagnac effect is neglected). The
Earth-based horizontal MIGO can thus measure $R_{0x0y}>4\times10^{-24}$
s$^{-2}$ in a one second integration time. With a local curvature of
$1.23\times10^{-6}$ s$^{-2}$, it will certainly be possible to measure the
Earth's Riemann curvature tensor if the contribution from the Sagnac effect
can be subtracted from the signal independent of the Earth's curvature's. It
will also be able to measure the Riemann curvature tensor of the Sun
($|R_{0i0j}|=3.2\times10^{-14}$ s$^{-2}$), and the Moon ($|R_{0i0j}
|=7.0\times10^{-14}$ s$^{-2}$) with MIGO; both curvatures vary predictably
with time, and their signal can be discriminated from the static phase shifts
cause by the Sagnac effect and the Earth. For comparison, a GW with an strain
amplitude of $10^{-21}$, and a frequency of $10^{4}$ Hz from a supernova
source shown in figure $\ref{h-Burst}$ has a rms $|R_{0i0j}|\sim2.5\times
10^{-14}$ s$^{-2}$, similar to the nearly static Riemann curvature of the Sun
and Moon. If it is possible to subtract out both the Sagnac and Earth's
curvature contributions, at a $|R_{0i0j}|\sim1\times10^{-17}$ s$^{-2}$, the
Earth-based horizontal MIGO will be able to measure the Lense-Thirring field
of the Earth as well.

On a classical level, this Riemann-curvature-driven effect on the atom
is surprising. In general relativity, an atom
is modelled as a spin-zero, point-like test particle, and thus
should not be directly affected by the curvature. However, on the
quantum level the atom is described by a \textit{delocalized}
wavefunction $\psi$, which is a section of a $U(1)$ line bundle
over the local spacetime manifold. Curvature-driven effects of the
$U(1)$ fibers of the line-bundle above a \textit{flat base
manifold} is well known in quantum mechanics, and is the
fundamental cause of Berry's phase \cite{Berry, Chiao}. It should
not be surprising that the Riemann curvature of a \textit{curved base
manifold} will cause phase shifts as well, in a gravitational 
version of the Aharonov-Bohm effect \cite{GLF}.

\subsubsection{Spatial Variations of GWs}

The length of LIGO's arms are such that $L_{LIGO}
\le\lambda_{min}/2\pi$, where $\lambda_{min}$ is the shortest
wavelength of GWs in LIGO's frequency range. Variations in the GW
in space can then be neglected, and only variations in time
matter, leading to Eq.~$(\ref{geodesicdeviation})$. However, if
the interferometer is \textit{larger} than wavelength of the GW,
general relativity predicts that additional terms that depend on
the velocity of the particle as well as the \textit{spatial}
variation of the GW\textemdash the magnetic part of the Weyl
tensor of the GW\textemdash will now appear in
Eq.~$(\ref{geodesicdeviation})$. It is in principle possible to
measure these new effects with MIGO; at $17$ km, the length of the
Space-based MIGO is comparable to the wavelength of a GW with a
frequency of roughly 3 kHz. As can be seen in figure
$\ref{h-Burst}$, there are a number of GW sources above this
frequency that can potentially be used to probe these effects.

\section{Conclusions}

In the first part of this paper, we have explored the implications, broadly
speaking, of non-Euclidean geometry for quantum physics. The problem of the
interaction of gravitational radiation with quantum matter, including
nonrelativistic quantum many-body-systems, was explored. In the second part of
this paper, we have explored in detail one particular aspect of this problem,
namely, the detection of astrophysical sources of the GWs by means of
matter-wave interferometry (MIGO). Our theoretical analysis indicates that in
terms of sensitivity, size, capability, and flexibility, the \textit{quantum} methods
embodied in MIGO have overwhelming advantages over the \textit{classical} methods
embodied in LIGO in studying general relativistic effects. What before took
many kilometers can now be done in a couple hundred meters, and what
took millions of kilometers can now be done with only thousands of
meters. The impact of MIGO will go beyond simply being another
astronomical instrument, however. With the advantage in MIGO that
slowly moving atoms are predicted to have over ultrarelativistic ones, it
will also spur theoretical and experimental studies that probe the
intersection of \textit{nonrelativistic} quantum mechanics with general
relativity \cite{Chiao}. As an example, we note that limitations to
MIGO's sensitivity are, at their most fundamental level, imposed by
quantum mechanics, and thus cannot be exceeded. Since MIGO in its
essence measures local curvature, like Wigner
\cite{Wigner1, Wigner2}, we ask: What are the fundamental quantum
limitations to the measurement of the Riemann curvature tensor arising
from the uncertainty principle? This leads naturally to the age-old
question, approached now from a different angle: What does it imply
for Einstein's theory of general relativity, and for geometry as a
whole, if curvature cannot be \textit{measured} beyond these fundamental
quantum limits? This question, which, in order to answer, was once
thought to require energy scales on the order of the Planck
energy\textemdash the energy where general relativity and the other
forces of nature are expected unify\textemdash can now be addressed
using matter-wave interferometry, once the fundamental quantum limits
to MIGO's sensitivity have been  established.

\section{\noindent{\textbf{Acknowledgements:}}}

\noindent ADS and RYC were supported by a grant from the Office of Naval
Research. We thank Peter Bender, John Garrison, Theodore H\"ansch, Jon Magne
Leinaas, Andrew P.~Mackenzie, Richard Marrus, Chris McKee, Joseph Orenstein,
William D. Phillips, Marlan O.~Scully, Jan Peter Toennies, and Rainer
Weiss for many clarifying and insightful discussions.

\appendix{}

\section{Brief Review of Gravitational Waves}

In this appendix, we present a brief review of GWs in the linearized
gravity limit; the reader is referred to \cite{MTW} or \cite{Wald}
for a complete review.

Given a spacetime manifold $\mathcal{M}$, the metric $g_{\mu\nu}$
gives the measure of length, $ds^2=g_{\mu\nu} dx^\mu dx^\nu$, on
$\mathcal{M}$. The precise form of $g_{\mu\nu}$ depends on the
coordinate system chosen, and if $\tilde{x}_\mu$ is another choice of
coordinates, then the coordinates $x_\mu$ expressed in the coordinates
$\tilde{x}_\mu$ is simply $x_\mu(\tilde{x})$. Since the length of a vector in
$\mathcal{M}$ cannot depend on the choice of coordinate systems,
$g_{\mu\nu}(x) dx^\mu dx^\nu = g_{\mu\nu}(\tilde{x}) d\tilde{x}^\mu
d\tilde{x}^\nu$, and by using the chain rule, we see that in a
coordinate transformation the metric changes by
\begin{equation}
g_{\mu\nu}(\tilde{x})= g_{\alpha\beta}(x)\frac{\partial x^\alpha}{\partial
  \tilde{x}^\mu}\frac{\partial x^\beta}{\partial \tilde{x}^\nu}.
\label{CoorTrans}
\end{equation}

No physically measurable quantities can depend on the choice of
coordinates. In Newtonian gravity, and in special relativity, we
require physics to be invariant under global coordinate
transformations. In general relativity, we require the theory to
be invariant under \textit{general} coordinate transformations.
Consequently, the choice of coordinate systems is often called a
`gauge choice', in analogy to EM, and general
coordinate-transformation-invariance is often referred to as `gauge
invariance' \cite{Yang}.

We are interested in the propagation of GWs in the spacetime
surrounding the Earth. Because of the relatively small mass
of the Earth, the spacetime can be approximated as
being flat. If $\eta_{\mu\nu}=diag(-1,1,1,1)$ is the Minkowski metric,
we can approximate $g_{\mu\nu}\approx \eta_{\mu\nu} +
h_{\mu\nu}$, where the components of $h_{\mu\nu}$ are
`small' compared with unity, and $h_{\mu\nu}$ represents the `strains'
in the flat spacetime caused by the passage of a GW. It is treated as
just another tensor on $\mathcal{M}$, and its indices are raised and
lowered using $\eta_{\mu\nu}$: $h_\mu^\nu = \eta^{\nu\alpha}
h_{\alpha\mu}$. In linearized gravity, we keep terms only linear in
$h_{\mu\nu}$; since $g^{\mu\alpha}g_{\alpha\nu} = \delta_\nu^\mu$,
$g^{\mu\nu} \approx \eta^{\mu\nu}- h^{\mu\nu}$.

Even though we are working in the linearized gravity limit, there is
still a remnant of the coordinate transformation invariance
Eq.~$(\ref{CoorTrans})$ left. If $x_\mu = \tilde{x}_\mu+\xi_\mu$ where
$\xi_\mu$ is a `small', arbitrary vector, then from
Eq.~$(\ref{CoorTrans})$, $h_{\mu\nu}$ transforms to
$\tilde{h}_{\mu\nu} = h_{\mu\nu}+\partial_\mu\xi_\nu +
\partial_\nu\xi_\mu$. The similarity between the transformation of
$h_{\mu\nu}$ under a \textit{coordinate} transformation, and the
transformation of the vector potential $A_\mu \to A_\mu +\partial_\mu
\phi$ under a $U(1)$ \textit{gauge} transformation is readily
apparent. Being a symmetric, $2\times 2$ tensor, $h_{\mu\nu}$ has ten
components. We have the freedom, as with the vector potential 
in EM, to make specific choices for $\xi_\mu$ and its
derivatives\textemdash eight variables in all\textemdash that will
determine our coordinate system. Thus, $h_{\mu\nu}$ contains only
$10-8=2$ physical degrees of freedom, or polarization states. The
usual coordinate choice is the TT gauge: $\eta^{\mu\nu}h_{\mu\nu}=0$,
$\partial^\mu h_{\mu\nu} = 0$, and, for GWs in Minkowski space, the
condition $h_{0\mu} =0$ is automatically follows.

In the linearized gravity limit, the Levi-Civita connection is in
general,
\begin{equation}
\Gamma^\alpha_{\mu\nu} =
            \frac{1}{2}\left(\partial_\mu h_\nu^\alpha
+ \partial_\nu h^\alpha_\mu -\partial^\alpha h_{\mu\nu}\right),
\label{Connection}
\end{equation}
where $\partial_\mu\equiv \partial/\partial x^\mu$, while the
Riemann curvature tensor and the Ricci tensor are, respectively
\begin{eqnarray}
R^\mu_{\>\>\>\>\nu\alpha\beta} &=& \frac{1}{2}
                \left\{
                \partial_\nu(
                    \partial_\alpha h_\beta^\mu
                    -
                    \partial_\beta h_\alpha^\mu)
                +
                \partial^\mu
                (\partial_\beta h_{\nu\alpha}
                -
                \partial_\alpha h_{\nu\beta})
                \right\},
\nonumber \\
R_{\mu\nu} &=& -\frac{1}{2}\Box h_{\mu\nu} + \frac{1}{2}
        \partial_\mu\partial^\alpha h_{\alpha\nu} +\frac{1}{2}
        \partial_\nu\partial^\alpha h_{\alpha\mu} - \frac{1}{2}
        \partial_\mu\partial_\nu h_\alpha^\alpha,
\label{Riemann}
\end{eqnarray}
were $R_{\mu\nu}=R^\alpha_{\mu\alpha\mu}$. Consequently, $R \equiv
\eta^{\mu\nu}R_{\mu\nu}= -\Box h_\mu^\mu + \partial_\mu \partial_\nu
h^{\mu\nu}$. For GWs in the TT gauge, the above equations reduce even
further, and we see that $R=0$, $R_{0\mu}=0$, and $R_{0i0j} =
-\ddot{h}_{ij}/2$.

The evolution equation for GWs comes from Einstein's equation,
\begin{equation}
R_{\mu\nu} -\frac{1}{2} g_{\mu\nu} R =\frac{8\pi G}{c^4} T_{\mu\nu},
\label{Einstein}
\end{equation}
where $G$ is Newton's constant, and $T_{\mu\nu}$ is the
 energy-momentum tensor of matter in the spacetime. Away from all
 sources, $T_{\mu\nu}=0$, and since $R=0$, we find
 that $R_{\mu\nu}=0$, or from Eq.~$(\ref{Riemann})$,
 $\partial^\alpha\partial_\alpha h_{ij}=0$, with the additional
 transversality condition $\partial^i h_{ij} =0$. Thus, like an EM
 wave in the  radiation gauge, $h_{ij}$ in the TT gauge is a
 transverse wave that satisfies the wave equation. It also can be
 expanded in plane waves,
\begin{equation}
h_{ij} = \sum_{s=+,\times} \sum_{k_xk_yk_z} \epsilon^{(s)}_{ij}(k)
 h_s(k) e^{i(\vec k\cdot \vec x - \omega t)},
\end{equation}
were $\epsilon^{(s)}_{ij}(k)$ are the unit polarization vectors, and the
 transversality condition becomes $k^i
 \epsilon^{(s)}_{ij}(k)=0$. Thus, in the plane perpendicular to $\vec
 k$, $\epsilon^{(s)}_{ij}$ can be represented by $2\times 2$ matrices,
where 
\begin{displaymath}
\epsilon^{(+)} =
\left(
\begin{array}{cc}
1&0\\
0&-1\\
\end{array}
\right),
\qquad
\epsilon^{(\times)} =
\left(
\begin{array}{cc}
0&1\\
1&0\\
\end{array}
\right).
\end{displaymath}
Sketches of the force-field lines produced by these polarizations are drawn
in figure $\ref{GW-Polarizations}$.

\section{Calculation of MIGO Phase Shifts}

We consider first the horizontal MIGO configuration figure
$\ref{H-V-MIGO}$.
Assuming that the atoms in figure $\ref{H-V-MIGO}$ are fast moving, in
the absence of a GW the atoms travel along the straight-line paths as
drawn. At time $t_r$, the $r$th atom is diffracted by the beam
splitter, and for the $n=+1$ order, the path that the atom travels is:
$v_{0y}=v_{\bot}$, $y_{0}(t)=v_{\bot}(t-t_r)$, and $v_{0x}=v_{\Vert}$,
$x_{0}(t)=v_{\Vert}(t-t_r)$, for $t_r<t<t_r+T/2$, where $T/2$ is the
time for the atom to travel from the beam splitter to the mirror. For
$t_r+T/2<t<t_r+T$ the path is: $v_{0y}=-v_{\bot}$, $y_{0}(t)=L_{\bot
}-v_{\bot}(t-t_r)$, while $v_{0x}=v_{\Vert}$ and
$x_{0}(t)=v_{\Vert}(t-t_r)$ still. Clearly, $L_{\bot}=v_{\bot}T$ and
$L_{\Vert}=v_{\Vert}T$. The velocities $v_{\bot}$ and $v_{\Vert}$ are
the initial velocities of the atom as it leaves the
diffraction-grating beam splitter, and just like for light,
$v_{\bot}=2\pi\hbar/ma$ where $a$ is the periodicity of the
grating. Then $v_{\Vert}=(v_{s}^{2}-v_{\bot}^{2})^{1/2}$ where
$v_{s}$ is the velocity of the beam incident the beam
splitter. Gravity can be neglected as long as $v_{\Vert}T>>gT^{2}/2$
where $g$ is the acceleration due to Earth's gravity. Since our focus
is on GWs, we approximate $g$ as a constant, and neglect local
curvature effects from stationary sources such as the Earth. A similar
set of equations hold for the $n=-1$ order with
$v_{\bot}\rightarrow-v_{\bot}$.

If a GW with $h_{ij}(t)$ is present, the paths of the atom will be
slightly perturbed from straight lines, and we take $x=x_{0}+x_{1}$ and
$y=y_{0}+y_{1}$, where $x_{1}$ and $y_{1}$ are deviations from the
unperturbed paths $x_{0}$ and $y_{0}$. These perturbations satisfy the
geodesic deviation equations of motion in
Eq.~(\ref{geodesicdeviation}), but with $(x,y)$ approximated as
$(x_{0},y_{0})$ on the right hand side.

To take into account the action of the GW on the mirrors, we model the
mirror and its connection to the frame of the interferometer as a
spring with a resonance frequency $f_{0}$\textemdash which depends on
the size of the interferometer as well as its material
properties\textemdash and a quality factor $Q$. We denote by $(X(t),
Y(t))$ the position of the mirror along the $n=+1$ path, and since in
the absence of a GW the mirror does not move, we take $X(t) = L_\|/2 +
X_1(t)$, and $Y(t) = L_\bot/2 + Y_1(t)$, where $X_1(t)$ and $Y_1(t)$
are perturbations of the mirror's position due to the GW. They are
also solutions of Eq.~$(\ref{geodesicdeviation})$, but because the
mirrors are connected to springs,
\begin{eqnarray}
\frac{d^2X_1}{dt^2} + \frac{2\pi f_0}{Q} \frac{dX_1}{dt} + (2\pi
f_0)^2 X_1(t)& =&
\frac{1}{2} \frac{d^2h_{xx}}{dt^2} \left(\frac{L_\|}{2}\right) +
\frac{1}{2} \frac{d^2h_{xy}}{dt^2} \left(\frac{L_\bot}{2}\right),
\noindent
\\
\frac{d^2Y_1}{dt^2} + \frac{2\pi f_0}{Q} \frac{dY_1}{dt} + (2\pi
f_0)^2 Y_1(t) &=&
\frac{1}{2} \frac{d^2h_{yy}}{dt^2} \left(\frac{L_\bot}{2}\right) +
\frac{1}{2} \frac{d^2h_{xy}}{dt^2} \left(\frac{L_\|}{2}\right),
\label{mirrors}
\end{eqnarray}
which are driven simple harmonic oscillators.

The action corresponding to the geodesic deviation equation is \cite{ADS1995}
\begin{equation}
S=\int_{t_r}^{t_r+T}dt\left(\frac{1}{2}m\vec{v}^{2}-
\frac{1}{2}m\dot{h}_{ij}v^{i}x^{j}\right),
\label{action2}
\end{equation}
Writing $\vec{v}=\vec{v}_{0}+\vec{v}_{1}$, we find that for the $n=+1$ path
the change in the action due to the GW is
\begin{align}
\frac{S_{+1}}{m}
\approx &\> L_{\Vert}\Bigg(  v_{1x}(t_r+T)-\frac{1}{2}L_{\Vert}\dot{h}
_{xx}(t_r+T)+\frac{1}{2}v_{\Vert}h_{xx}(t_r+T)-\frac{1}{2}v_{\bot}
\left[h_{xy}(t_r+T)-h_{xy}(t_r+T/2)\right]\Bigg)+
\nonumber\\
&  {}\frac{L_{\bot}}{2}\left[  v_{1y}(t_r+T^{-}/2)-v_{1y}(t_r
+T^{+}/2)\right]  +\frac{1}{2}v_{\bot}L_{\bot}h_{yy}(t_r+T/2)-
\nonumber\\
&  {}\frac{1}{2}\int_{t_r}^{t_r+T}\left(v_{\Vert}^{2}h_{xx}
(t)+v_{\bot}^{2}h_{yy}(t)\right)dt- v_{\bot}v_{\Vert}\left\{  \int_{t_r
}^{t_r+T/2}h_{xy}(t)\>dt-\int_{t_r+T/2}^{t_r+T}h_{xy}(t)\>dt\right\},
\label{S-n+1}
\end{align}
after successive integration by parts of Eq.~$(\ref{action2})$, and using
Eq.~$(\ref{geodesicdeviation})$. The velocities $v_{1y}(t_r+T^{-}/2)$
and $v_{1y}(t_r+T^{+}/2)$ are the perturbed velocities of the atom
right before and right after the mirrors, respectively. Since the
passage of the GW will shift the velocity of the mirrors as well as
the atoms, a jump condition of the $y$-velocity at the mirror
requires that
\begin{equation}
v_{1y}(t_r+T^{+}/2)=-v_{1y}(t_r+T^{-}/2)+V_{1y}(t_r+T/2).
\end{equation}
Taking the Fourier transform of $h_{ij}$, and neglecting the
transients in the solution of Eq,~$(\ref{mirrors})$, we find for a
horizontal MIGO that
\begin{equation}
\Delta\phi_{MIGO}^{hor}(f)=-\frac{m}{\hbar}\pi
L_{\bot}L_{\Vert}ifh_{\times}(f)e^{-i\pi fT}F_{h}(fT),
\label{MIGO-H-V-Phase}
\end{equation}
for a GW with amplitude $h_{\times}(f)$ and frequency $f$. Only the
$\times$ polarization causes a net phase shift in the interferometer;
the $+$ polarization does not, as can be seen due to the inherent
bilateral symmetry of the interferometer. The horizontal resonance
function is
\begin{equation}
F_{h}(fT)=1-2e^{i\pi fT/2}\hbox{sinc}\left(\frac{\pi fT}{2}\right)+
\left[\hbox{sinc}\left(\frac{\pi fT}{2}\right)\right]^{2}-
\frac{1}{2}\frac{f^{2}}{f^{2}-f_{0}^{2}+if_{0} f/Q},
\label{F_h}
\end{equation}
 where $\hbox{sinc}(x)\equiv\sin{x}/x$. The analysis of the vertical
 configuration MIGO follows in the same way. The overall form of
 Eq.~$(\ref{MIGO-H-V-Phase})$ still holds, with only the replacement
 $F_{h}(fT/4)$ by $2 F_{v}(fT)$, where now
\begin{equation}
F_{v}(fT)= 1-\left[\hbox{sinc}\left(\frac{\pi
    fT}{2}\right)\right]^{2} + \frac{i}{\pi f T/2}
\left[1-\hbox{sinc}(\pi
    fT)\right]-\frac{1}{2}\frac{f^{2}}{f^{2}-f_{0}^{2}+if_{0} f/Q},
\label{F_v}
\end{equation}
for this configuration.

If we consider the low frequency $f\to0$ limit of Eqs.~$(\ref{F_h})$
and $(\ref{F_v})$, we find that $F_h(fT) \approx -i\pi fT$, while
$F_v\sim i\pi fT/3$, irrespective of how the mirror is mounted onto the
frame of the interferometer. This is not the case in the high
frequency limit. We find that in the infinitely-rigid interferometer
limit where $f_0\to\infty$, and then $f\to\infty$, both resonance
factors approach one. If, however, the mirrors are freely suspended
like those of LIGO, then $f_0\to0$ first. Then when the $f\to
\infty$ limit is taken, both resonance functions approach one-half. In
the MIGO graphs of figure $\ref{h-Burst}$, we took as the speed of
sound for the material used to construct the interferometer frame as
$6000$ m/s, and that the motion of the mirrors were critically damped with
$Q=1$. This resulted in the small dips at the high-frequency end of
the MIGO graphs.

\section{Effects of Thermal Noise on MIGO}

We consider the mirrors to be rigidly attached to the frame of the
interferometer. (This analysis can be repeated with the mirrors attached to
springs as described in \textbf{Appendix B}). The finite temperature of the
mirror induces fluctuations on the surface of the mirror, i.~e., transverse
surface sound waves, which will induce random Doppler-like shifts in the
velocity of the atoms after it hits the mirror. Such velocity shifts are not
important for LIGO because the speed of light is much larger compared to the
velocity of these fluctuations. For slowly moving atoms, on the other
hand, these fluctuations are important, and can potentially degrade
the sensitivity of MIGO. 

We model the surface of the mirror as a continuous sheet with transverse
surface fluctuations $u(\vec{x}, t)=u_{0} \sin(\vec{k}_{s}\cdot\vec{x} -
\omega_{s} t+\phi_{s})$, where $\vec{x}=(x,y)$ is a vector at the surface of
the mirror, $\vec{k}_{s}$ is the wavenumber of the sound mode, $\omega_{s}$ is
its frequency, and $\phi_{s}$ is its random phase. The dispersion relation for
the sound wave is $\omega_{s} = v_{sound} k_{s}$ where $v_{sound}$ is the
transverse speed of sound for the mirror. The surface of the mirror in the
presence of the sound mode is defined by $z=u(\vec{x}, t)$.

Consider an atom with wavefunction $\psi_{inc}(\vec{x},z,t)=A_{0}
\exp\{i(k_{\bot}z+\omega_{\bot}t)\}$ normally incident on the mirror. The
reflected wavefunction $\psi_{ref}(\vec{x},z,t)$ is determined by the
Schr\"{o}dinger equation, and the boundary conditions that the total
wavefunction vanishes at $z=u(\vec{x},t)$. The total current \textit{normal to
the surface of the mirror} vanishes there as well. Because this surface is
changing with time, this is a moving boundary problem that cannot be solved
exactly. We note, however, that for $\omega_{s}>>\omega_{\bot}$, the surface
of the mirror oscillates many times during the time when the atom reflects off
of it; these oscillations cancel each other, and do not adversely affect the
motion of the atom. On the other hand, when $\omega_{s}<<\omega_{\bot}$, the
surface is frozen in time, and the problem reduces to an atom reflecting off a
rough surface. It is straightforward to see that for this case any variations
in the reflected atom's velocity normal to the mirror is due to changes in the
direction of the normal vector to the surface of the mirror. This causes a
velocity shift of the atom that is proportional to $(k_{s}u_{0})^{4}$, and as
we shall see, is not the dominant effect.

Consider now the case when $\omega_{s} \approx\omega_{\bot}$. Expanding
$\psi_{inc}$ at the surface of the mirror, we get
\begin{eqnarray}
\psi_{ref}\left[\vec{x}, z=u(\vec{x},t), t\right]&=& -A_{0}
\exp\{i\left( k_{\bot}u(\vec{x},t)+\omega_{\bot}t\right)\},
\nonumber
\\
&=& -A_{0}\sum_{q=-\infty}^{q=\infty} J_{q}(k_{\bot}u_{0})
\exp\left\{i\left(\omega_{\bot}- q\left[\vec k_\bot\cdot \vec
  x-\omega_{s}t+\phi_s\right]\right)\right\}.
\end{eqnarray}
The mirror surface acts as a FM modulator, and produces sidebands with an
initial amplitude given by the $q$th order Bessel function
$J_{q}(k_{\bot}u_{0})$. The amplitude of the mode $u_{0}=
v_{th}/\omega_{s}$, where $v_{th}= (k_{B}T_{mirr}/M)^{1/2}$ with
$T_{mirr}$ the temperature of the mirror with mass $M$, while
$k_{\bot}= 2\pi/a$, where $a$ is the periodicity of the diffraction
grating used in the beam splitter. The product $k_{\bot}u_{0}=
k_{\bot}v_{th} /\omega_{s}\approx k_{\bot}v_{th}/\omega_{\bot}=
v_{th}/v_{\bot}<<1$. Consequently, only the $q=1$ sideband is
important (since $\omega_{\bot}\approx\omega_{s}$, the $q=-1$ sideband
is not a propagating mode).

Thus, since $J_{0}(k_{\bot}u_{0})\approx1$, the incident wavefunction is
reflected backward with almost the same amplitude it came in with, but now an
additional $q=1$ component with amplitude $J_{1}(k_{\bot}u_{0})$ mixed in. But
because the phase of the sound mode is arbitrary, the wavenumber of this
reflected wave is approximately
$-k_\bot+k_{1}(v_{th}/v_{\bot})^{2}/2$, where
$k_1=(4m\omega_{\bot}/\hbar)^{1/2}$ is the wavenumber of the $q=1$
reflected wavefunction.  The linear term in the expansion of
$J_{1}(k_{\bot}u_{0})$ vanishes since roughly half the time the sound
mode is in phase with the incident atom, and half the time it is out
of phase. Thus, the fractional change in the atom's velocity is due to
thermal motion of the mirror is $\Delta
v/v_{\bot}=(v_{th}/v_{\bot})^{2}$, and is independent of the material
properties of the mirror, as expected. This shift in the atom's
velocity introduces a thermal-noise-limited sensitivity
\begin{equation}
\tilde{h}(f)_{th}=\frac{L_{\bot}\Delta v}{2\pi A|F(fT)|f^{3/2}},
\end{equation}
which dies out faster than the shot-noise-limited sensitivity $\tilde
{h}(f)_{shot}\sim1/f$ for high frequencies, but it is dominant for low
frequencies. The thermal- and shot-noise-limited sensitivities are equal at
$f_{turn}=\dot{N}(mL_{\bot}\Delta v/\hbar)^{2}$. For a one-kilogram-mass
mirror at a temperature of 1 K with a $6000$ m/s transverse speed of sound,
this is roughly $10^{-7}-10^{-8}$ Hz for the $^{133}$Cs atom MIGOs
shown in figure $\ref{h-Burst}$, and causes the curves of the
Earth-based vertical MIGO and the Space-based MIGO to angle slightly
upwards at low frequencies. For the Earth-based horizontal MIGO this
frequency is roughly $10^{-12}$, and thermal effects are negligible.


\begin{thebibliography}{9}
\bibitem {Comment-1} This analogy is usually made between GWs and
  non-Abelian field theories such as $SU(2)$ \cite{Yang}. We will, bowever, be
  working in the linearized gravity limit where the non-linear aspects
  of GWs are neglected. In this limit, the analogy with EM holds well.
\bibitem {Comment} Weinberg showed in \cite{Weinberg} that when one
  attempts to generate a GW by means of an acoustical excitation of a
  Weber bar, the majority of the energy in the bar is dissipated as
  heat. Only a small fraction of this energy is converted into
  GWs. While valid for classical systems like the Weber bar, this
  analysis often fails for quantum many body systems such as the
  quantum Hall fluid, where the existence of a finite energy gap
  allows for macroscopic, dissipationless quantum flows in the fluid.
\bibitem {Comment2} While the sensitivity of LIGO can be, in
  principle, improved using squeezed light \cite{Caves}, the
  sensitivity of MIGO can be, in principle, improved by using squeezed
  atom states as well \cite{Meystre}. While the practicality of using
squeezed light is still in question, squeezed fermionic number states
are very robust.
\bibitem {Comment3} Stodolsky briefly mentioned the use of the
  synchronous coordinates as a justification for not considering the
  action of GWs with parts of the interferometer. This approach is in
  conflict with the usual coordinate system chosen to detect GWs.
\bibitem {Comment4} By which we mean that $v_{sound}/L_{charac}>> f$,
  where $L_{charac}$ is the characteristic length of the
  interferometer, and $v_{sound}$ is the speed of the sound of the
  material used to construct it.
\end{thebibliography}
\end{document}